\documentclass[preprint,12pt]{elsarticle}
\usepackage[margin=1.8cm]{geometry}
\usepackage{amsmath}
\usepackage{amsfonts}
\usepackage{amssymb}
\usepackage{graphicx}
\usepackage{lmodern}
\usepackage{hyperref}
\usepackage{dcolumn}
\usepackage{bm}
\usepackage[mathlines]{lineno}
\usepackage{float}
\usepackage[T1]{fontenc}
\usepackage{subfigure}
\usepackage{placeins}

\journal{Journal of Physics G}

\usepackage{numcompress}

\begin{document}

\begin{frontmatter}

\title{\textbf{Constrained Neutrino Mass Matrix and Majorana Phases}}



\author[mymainaddress]{Pralay Chakraborty}
\ead{pralay@gauhati.ac.in}

\author[mymainaddress]{Manash Dey}
\ead{manashdey@gauhati.ac.in}

\author[mymainaddress]{Subhankar Roy\corref{mycorrespondingauthor}}
\cortext[mycorrespondingauthor]{Corresponding author}
\ead{subhankar@gauhati.ac.in}

\address[mymainaddress]{Department of Physics, Gauhati University, India}

\begin{abstract}
We endeavour to constrain the neutrino mass matrix on the phenomenological ground and procure model-independent textures by emphasizing on the simple linear relationships among the mass matrix elements. These simple textures predict the two Majorana phases. In this regard, two types of parametrization of neutrino mass matrix: general and exponential are employed. We obtain fifty-three predictive neutrino mass matrix textures, out of which twenty-eight are associated with the general parametrization, and the rest belong to the exponential one. Apart from Type-A/P textures, the rest deal with the prediction of a few other oscillation parameters as well. We try to realize the proposed textures in the light of  $A_4$, $\Delta\,(27)$ and $T_7$ symmetry groups. 
\end{abstract}


\end{frontmatter}


\section{Introduction \label{Introduction}}

The Standard Model (SM) of Particle Physics framed in the light of $SU(3)_c\otimes SU(2)_L\otimes U(1)_Y$, is a renormalizable quantum field theory which depicts successfully the properties of the  fundamental particles and their interactions\,~\cite{a,b,c}. It incorporates the Higgs Mechanism which is responsible for the masses of all the fundamental particles and the gauge bosons participating in the weak interaction\,~\cite{d,e}. The Lagrangian of the theory involves one particular term named as Yukawa term which, after Higgs mechanism, yields the masses of the fundamental fermions.  This term involves both left handed and right handed components of the fermion field. Since the right handed neutrino is unobservable in the experiments, the SM excludes this possibility and thus the neutrino remains massless in the theory. But the phenomenon of neutrino oscillation\,~\cite{f1,f,g,h} vindicates the existence of nonzero  neutrino mass and hence the explanation of the neutrino mass takes us beyond the territory of SM (BSM). So that an Yukawa term for the neutrinos could be incorporated in the theory, either we have to posit the existence of right handed Dirac neutrino field, or we have to believe the neutrino to be a Majorana particle\,~\cite{i,j}. The Majorana nature is associated with the fact that the neutrino being a chargeless entity, resembles its own antiparticle state. If so, then we can incorporate  the Yukawa term for the Majorana neutrinos as shown below,

\begin{equation}
  \mathcal{L}_\nu = \nu^T_L C^{-1} M_\nu \nu_L+h.c,
  \label{equation 1}
 \end{equation}
   
where, $\nu_L$  denotes the left-handed neutrino field $\nu_L=(\nu_{eL}$, $\nu_{\mu L}$, $\nu_{\tau L})^T$ and $C$ is the  charge conjugation operator. The $M_\nu$ is the Majorana neutrino mass matrix\,~\cite{k,k1} which carries the information of neutrino mass eigenvalues. Because the neutrino is Majorana in nature, the $M_\nu$ is supposed to be complex symmetric. To understand the three masses, the $M_\nu$ has to be diagonalised and we have to move from the flavour basis ($\nu_L$) to the mass basis, $\nu_i = (\nu_1, \nu_2,\nu_3)$, where the $M_{\nu}$ is diagonal. Both basis systems are connected through a transformation as shown in the following,
                                                        
 \begin{equation}
 \begin{bmatrix}
 \nu_{eL}\\
 \nu_{\mu L}\\
 \nu_{\tau L}\\
 \end{bmatrix}
 =
 U
 \begin{bmatrix}
 \nu_1\\
 \nu_2\\
 \nu_3\\
 \end{bmatrix},
 \end{equation}

where, $U$ is a unitary matrix and it is defined as Pontecorvo-Maki-Nakagawa-Sakata (PMNS) matrix ~\cite{Maki:1962mu}. In general, it is defined in the basis where the charged lepton mass matrix is diagonal, and it diagonalizes the $M_\nu$ appearing in Eq.\,(\ref{equation 1})

 \begin{equation}
 M^{diag}_{\nu} = U^{\dagger} M_{\nu} U^*.
 \end{equation}
 
In the present work, we shall stick to the Majorana nature of neutrino. The Majorana nature of neutrino is the key to See-Saw mechanism\,~\cite{l1,l2,l3} which helps to understand the smallness of neutrino masses. It is needless to mention that the confirmation of the Majorana nature of neutrinos is subjected to the experimental evidences. 
 
The $M_\nu$ contains the information of all nine physical parameters within itself: three mass eigenvalues ($m_1, m_2, m_3$), three mixing angles (solar neutrino mixing angle $\theta_{12}$, reactor neutrino mixing angle $\theta_{13}$ and atmospheric neutrino mixing angle $\theta_{23}$) and three phases ($\delta$, $\alpha$, and $\beta$) where, $\delta$ is the Dirac type CP violating phase whereas, $\alpha$ and $\beta$ resemble the two Majorana phases. The PMNS matrix is observable in the experiments. However, the oscillation experiments do not provide any information regarding the two Majorana phases. Apart from that, the three neutrino mass eigenvalues are still in the dark as the oscillation experiments give the information of $\Delta m_{21}^2 = m_2^2-m_1^2$ and absolute value of $\Delta m_{31}^2=m_3^2-m_1^2$. We understand that the neutrino mass ordering follow the pattern: either $m_1< m_2< m_3$ (Normal ordering) or $m_2>m_1>m_3$ (Inverted ordering). The ignorance of the absolute neutrino mass eigenvalues in turn makes it difficult to understand the Majorana phases. Although, we cannot measure the absolute neutrino mass eigenvalues, an upper bound on the sum of the three neutrino masses is prominent from the cosmological observation\,~\cite{m}. The upper bounds on the sum of the three neutrino masses are listed in Table \ref{table:upper bound of sum of neutrino masses}.

We highlight that the present work will stick to the normal ordering of neutrino masses. Moreover, the experiments have not clearly indicated whether $\theta_{23}> 45^{\circ}$ or $\theta_{23}<45^{\circ}$. The details of the physical oscillation parameters\,~\cite{n} are highlighted in Table \ref{table:Values Of Parameters in NO}.

If the neutrinoless double beta decay ($0\nu\beta\beta$ decay)\,~\cite{o1,o2,o3,o4}, which violates the lepton number conservation, were discovered  then it would directly justify the Majorana nature of neutrino. In addition to the dependence on the phase space $G^{0\nu}$ and the matrix element $M^{0\nu}$, the $0 \nu \beta \beta$ decay rate $\Gamma$ is found to be proportional to effective Majorana mass $m_{\beta\beta}$,
 
 \begin{equation}
 \Gamma\sim G^{0\nu}. M^{0\nu}.m^2_{\beta\beta}.
 \end{equation}

It is to be noted that the only parameter which is found to be feasible from  $0 \nu \beta \beta$  decay is the effective Majorana neutrino mass: \[m_{\beta \beta}=|\sum_{k=1}^{3}{U^2_{1k}m_k}|.\] where, $m_1$, $m_2$ and $m_3$ are the three mass eigenvalues, $U_{11}$,$U_{12}$ and $U_{13}$ are the elements of the PMNS matrix which contains the information of the Majorana phases $\alpha$ and $\beta$. Several experiments provides the upper bounds of $m_{\beta \beta}$: SuperNEMO(Se$^{82}$) as $67-131$ meV,  GERDA(Ge$^{76}$) as $104-228$ meV, EXO-200(Xe$^{136}$) as $111-477$ meV, CUORE(Te$^{130}$) as $75-350$ meV and KamLAND-Zen(Xe$^{136}$) as $61-165$ meV ~\cite{p, Agostini:2022zub, CUORE:2019yfd, GERDA:2019ivs, KamLAND-Zen:2016pfg,p1,p2,p3,p4}. However, in the upcoming years, the experiments aim to constrain the upper bounds of $m_{\beta\beta}$ further.

The Majorana hypothesis has got many theoretical advantages. If the neutrino were a Majorana particle, then we would not require the right handed neutrino in the SM content. With the $\nu_L$ and its Charge conjugate, $\nu_L^C$, the Majorana mass term  $\sim\, m\, \bar{\nu}^C_L\nu_{L}$ can be included in the framework, where, $m$ is the mass of the Majorana neutrinos. However, if we try to look into the origin of neutrino mass via see-saw mechanism which is expected to occur at a higher energy scale ($\approx 10^{14}$\,GeV), the introduction of heavy right-handed neutrino is inevitable. Lastly, we see that the  Majorana fermions are permitted by the Poincar\'e group, and they constitute the simplest spinorial representation\,~\cite{q}. In our work, we shall try to work out the promising textures of $M_\nu$ predicting the Majorana phases. The plan of the paper is mentioned as shown: In section \ref{Parametrization}, we give the parametrization of PMNS matrix and Majorana neutrino mass matrix. In section \ref{Formalism}, we illustrate the formalism of our work. The symmetry realization of two specific textures is added in section \ref{Symmetry Realization}. We write the summary and discussion of our work in section \ref{Summary}.

\section{Parametrization of PMNS Matrix and Majorana Neutrino Mass Matrix} \label{Parametrization}

The PMNS matrix appears in the weak interaction Lagrangian ($\mathcal{L}_{int}$) of the SM and the latter can be expressed as in the following,

\begin{eqnarray}
\mathcal{L}_{int}&=& \frac{g}{\sqrt{2}} W^{-}_\mu \begin{bmatrix}
                      \bar e_L  \bar \mu_L  \bar \tau_L
                      \end{bmatrix} \gamma^\mu U \begin{bmatrix}
                      \nu_1\\
                      \nu_2\\
                      \nu_3\\
                      \end{bmatrix},                       
\label{Interaction Lagrangian}
\end{eqnarray}

where, $U$ is the PMNS matrix. In the above expression, the ($e_L, \mu_L, \tau_L$) represents basis where the charged lepton mass matrix, $M_{l}$ is diagonal. If somehow, $M_{l}$ is not diagonal, then we have to diagonalise it, $M_{l}^{diag}= U_{lL}^{\dagger}.M_{l}U_{lR}$, where, $U_{lL}$ and $U_{lR}$ are the left handed and right handed diagonalizing matrices respectively. Under this situation, the PMNS matrix is defined as $U= U^\dagger_{lL} U_\nu$, where $U_{\nu}$ is the diagonalizing matrix for $M_{\nu}$. The $U$, being a $3\times 3$ unitary matrix, requires  nine physical parameters; three angles and six phases. The three unphysical phases out of six, can be absorbed by redefining the left handed charged lepton fields. If the neutrino were a Dirac particle, then the other two phases could be absorbed by the redefinition of the neutrino mass eigenstates appearing in the interaction Lagrangian [See Eq.\,(\ref{Interaction Lagrangian})]. Thus, for Dirac neutrino, $U$ is parametrised by three angles and one phase. But if we consider the neutrino as Majorana particle, then these two phases will again appear in the Majorana mass term via $\nu_L$ appearing in Eq.\,(\ref{equation 1}). This stipulates that we need three angles and three phases to parametrize $U$. The Particle Data Group adopted a parametrization for $U$, known as the Standard Parametrization where $U$ is parametrized by three mixing angles ($\theta_{12}$, $\theta_{13}$ and $\theta_{23}$) and three phases ($\alpha$, $\beta$ and $\delta$) as shown below,

\begin{equation}
   U = \begin{bmatrix}
c_{12}c_{13} & c_{13}s_{12} & s_{13}e^{-i\delta}\\
-c_{23}s_{12}-c_{12}s_{13}s_{s23}e^{i\delta} & c_{12}c_{23}-s_{12}s_{13}s_{23}e^{i\delta} & c_{13}s_{23}\\
s_{12}s_{23}-c_{12}c_{23}s_{13}e^{i\delta} & -c_{12}s_{23}-s_{12}s_{13}c_{23}e^{i\delta} & c_{13}c_{23}\\
\end{bmatrix} \begin{bmatrix}
e^{i\alpha} & 0 & 0\\
0 & e^{i\beta} & 0\\
0 & 0 & 1\\
\end{bmatrix}
\end{equation}     

where, $c_{ij}=\cos \theta_{ij}$ and $s_{ij}=\sin \theta_{ij}$.

We represent the $M_{\nu}$ in the following manner,

\begin{equation}
M_\nu = \begin{bmatrix}
m_{ee} & m_{e\mu} & m_{e\tau}\\
m_{e\mu} & m_{\mu\mu} & m_{\mu\tau}\\
m_{e\tau} & m_{\mu\tau} & m_{\tau\tau}\\
\end{bmatrix},
\end{equation}

where,

\begin{eqnarray}
m_{ee} &=&\,m_1\,c^2_{12}\,c^2_{13}\,e^{2i\alpha}\,+\, m_2\,c^2_{13}\,s^2_{12}\,e^{2i\beta}\,+\,m_3\,s^2_{13}\,e^{-2i\delta}\label{equation 7},\\
m_{e\mu} &=&\,m_2\,s_{12}\,c_{13}\,e^{2 i \beta }\,(c_{12}\, c_{23}\, -\,s_{12}\,s_{13}\,s_{23}\,e^{i \delta })\,+\, m_3\,s_{13}\,c_{13}\,s_{23}\,e^{-i \delta }\,-\,m_1\,c_{12}\,c_{13}\,\nonumber\\&&e^{2 i \alpha }\,(\,s_{12} \,c_{23}\,+\,c_{12}\,s_{13}\,s_{23}\,e^{i \delta }),\\
m_{e\tau}&=&\,m_1\,c_{12}\,c_{13}\,e^{2 i \alpha }\,(s_{12}\, s_{23}\,-\,c_{12}\,s_{13}\,c_{23}\,e^{i \delta })\,+
\,m_3\,s_{13}\,c_{13}\,c_{23}\,e^{-i \delta}\,-
\,m_2\,s_{12}\,c_{13}\,\nonumber\\&&e^{2 i \beta }\,
(c_{12}\,s_{23}\,+\,s_{12}\,s_{13}\,c_{23}\,e^{i \delta }),\\
m_{\mu \mu}&=&\,m_1\,e^{2 i \alpha}\,(s_{12}\,c_{23}\,+\,c_{12}\,s_{13}\,s_{23}\,e^{i \delta}\,)^2\,+\,m_3\,c^2_{13}\,  s^2_{23}\,+\,m_2\,e^{2 i \beta}\,(c_{12}\,c_{23}\,-\,s_{12}\nonumber\\&&\,s_{13}\,s_{23}\,e^{i \delta})^2,\\
m_{\mu \tau}&=&\,m_3\,c^2_{13}\, s_{23}\,c_{23}\,-\,m_1\,e^{2i\alpha}\,(s_{12}\,s_{23}\,-\,c_{12}\,s_{13}\,c_{23}e^{i \delta})\,(s_{12}\,c_{23}\,+\,c_{12}\,s_{13}\,s_{23}\,e^{i \delta})\,\nonumber\\&&-\,m_2\,e^{2 i \beta}\,(c_{12}\,s_{23}\,+\,s_{12}\,s_{13}\,c_{23}e^{i \delta })\,(c_{12}\,c_{23}\,-\,s_{12}\,s_{13}\,s_{23}e^{i \delta }),\\
m_{\tau \tau}&=&\,m_1\,e^{2 i \alpha }\,(s_{12}\,s_{23}\,-\,c_{12}\,s_{13}\,c_{23}e^{i \delta})^2\,+\,m_2\,e^{2 i \beta}\,(\,c_{12}\,s_{23}\,+\,s_{12}\,s_{13}\,c_{23}e^{i \delta })^2\,+\,\nonumber\\&&m_3\,c^2_{13}\,c^2_{23}\label{equation 12}.
\end{eqnarray}

For simplicity, we normalize $M_\nu$ with respect to $m_3$ in our calculation. We shall use two different kinds of parametrization of neutrino mass matrix. 

Since a complex number $z$, can be expressed as $z=x\,+\,iy$ where $x$ and $y$ are two real numbers, hence the general neutrino mass matrix which is complex symmetric can be presented as shown below,

\begin{eqnarray}
M_\nu &=& \begin{bmatrix}
\text{Re}[m_{ee}] & \text{Re}[m_{e\mu}] & \text{Re}[m_{e\tau}]\\
\text{Re}[m_{e\mu}] & \text{Re}[m_{\mu\mu}] & \text{Re}[m_{\mu\tau}]\\
\text{Re}[m_{e\tau}] & \text{Re}[m_{\mu\tau}] & \text{Re}[m_{\tau\tau}]\\
\end{bmatrix} + i \begin{bmatrix}
\text{Im}[m_{ee}] & \text{Im}[m_{e\mu}] & \text{Im}[m_{e\tau}]\\
\text{Im}[m_{e\mu}] & \text{Im}[m_{\mu\mu}] & \text{Im}[m_{\mu\tau}]\\
\text{Im}[m_{e\tau}] & \text{Im}[m_{\mu\tau}] & \text{Im}[m_{\tau\tau}]\\
\end{bmatrix}
\label{general parametrization}
\end{eqnarray}

The other way of representing a complex number is $|z|\,e^{i\theta}$, where $\theta$ is the $arg\,[z]$. In this parametrization, the $M_{\nu}$ will appear as in the following,

\begin{equation}
M_\nu = \begin{bmatrix}
|m_{ee}|e^{i \text{arg}[m_{ee}]} & |m_{e\mu}|e^{i \text{arg}[m_{e\mu}]} & |m_{e\tau}|e^{i \text{arg}[m_{e\tau}]}\\
|m_{e\mu}|e^{i \text{arg}[m_{e\mu}]} & |m_{\mu\mu}|e^{i \text{arg}[m_{\mu\mu}]} & |m_{\mu\tau}|e^{i \text{arg}[m_{\mu\tau}]}\\
|m_{e\tau}|e^{i \text{arg}[m_{e\tau}]} & |m_{\mu\tau}|e^{i \text{arg}[m_{\mu\tau}]} & |m_{\tau\tau}|e^{i \text{arg}[m_{\tau\tau}]}
\end{bmatrix}
\label{exponential parametrization}
\end{equation}

Parametrizing the $M_\nu$ appearing in Eq.\,(\ref{general parametrization}) gives a liberty to visualize the real and imaginary parts of the matrix elements and thus makes it easier to draw some correlations among them. This motivates us to derive certain promising textures of the neutrino mass matrix, while the second parametrization emphasizes on the absolute value of the matrix element along with the argument of the latter. This parametrization is important for those phenomenological contexts where we try to look into the absolute version of the neutrino mass matrix to decipher certain textures. But to keep the discussion general and fruitful, we shall treat both $|m_{ij}|$ and $arg\,[m_{ij}]$ on equal footing. 
 
 We define the parametrization of $M_{\nu}$ appearing in Eq.\,(\ref{general parametrization}) as General Parametrization (GP), whereas that appearing in Eq.\,(\ref{exponential parametrization}) as Exponential Parametrization (EP).
 
\section{Formalism and Numerical Analysis}\label{Formalism}

As previously discussed, among nine parameters of $M_{\nu}$, three mass eigenvalues and the Majorana phases are not yet determined experimentally. Therefore, it is of significant merit to develop certain phenomenological frameworks that facilitate the prediction of the above mentioned observables. A predictive neutrino mass matrix inherently shelters certain correlations/constraints among its constituent elements. In this regard, several promising approaches are explored, such as texture zeros \cite{Ludl:2014axa}, hybrid textures\cite{Singh:2018bap} and vanishing minors\cite{Lashin:2009yd}, etc. However, deviating a little from the established methodologies, in the present work, we adopt a new approach emphasizing on simple linear correlation among the neutrino mass matrix elements.

Based on the parametrization discussed in the section (\ref{Parametrization}), we try to work out some specific textures of neutrino mass matrix which are consistent with the present data obtained from neutrino oscillation experiments. Moreover, we target to predict the two Majorana phases which are untouched in neutrino oscillation experiments. In this regard, three categories specified as Type - \textbf{A}, \textbf{B} and \textbf{C} are demonstrated in the light of GP [see Table \ref{table:Type-A textures}, \ref{table:Type-B textures} and \ref{table:Type-C textures}]. Similarly, based on EP, we design three categories of textures namely, Type - \textbf{P}, \textbf{Q} and \textbf{R} [see Table \ref{table:Type-P textures}, \ref{table:Type-Q textures} and \ref{table:Type-R textures}]. The Type - \textbf{A}/\textbf{P} emphasizes over the simultaneous existence of two constraint relations within $M_{\nu}$. On the other hand, Type - \textbf{B}/\textbf{Q} and \textbf{C}/\textbf{R} insist on the coexistence of three and four relations respectively. The Type-A/P family predicts two Majorana phases $\alpha$ and $\beta$; that B/Q predicts $\alpha$, $\beta$ and $\delta$ and Type-C/R predict $\theta_{23}$ in addition to $\alpha$, $\beta$ and $\delta$. The effective Majorana neutrino mass $m_{\beta\beta}$ is an observational parameter in neutrinoless double beta decay. In this regard, we attempt to the parameter $m_{\beta\beta}$ for all the proposed textures.

 \FloatBarrier
\begin{figure}[!]
  \centering
    \subfigure[]{\includegraphics[width=0.25\textwidth]{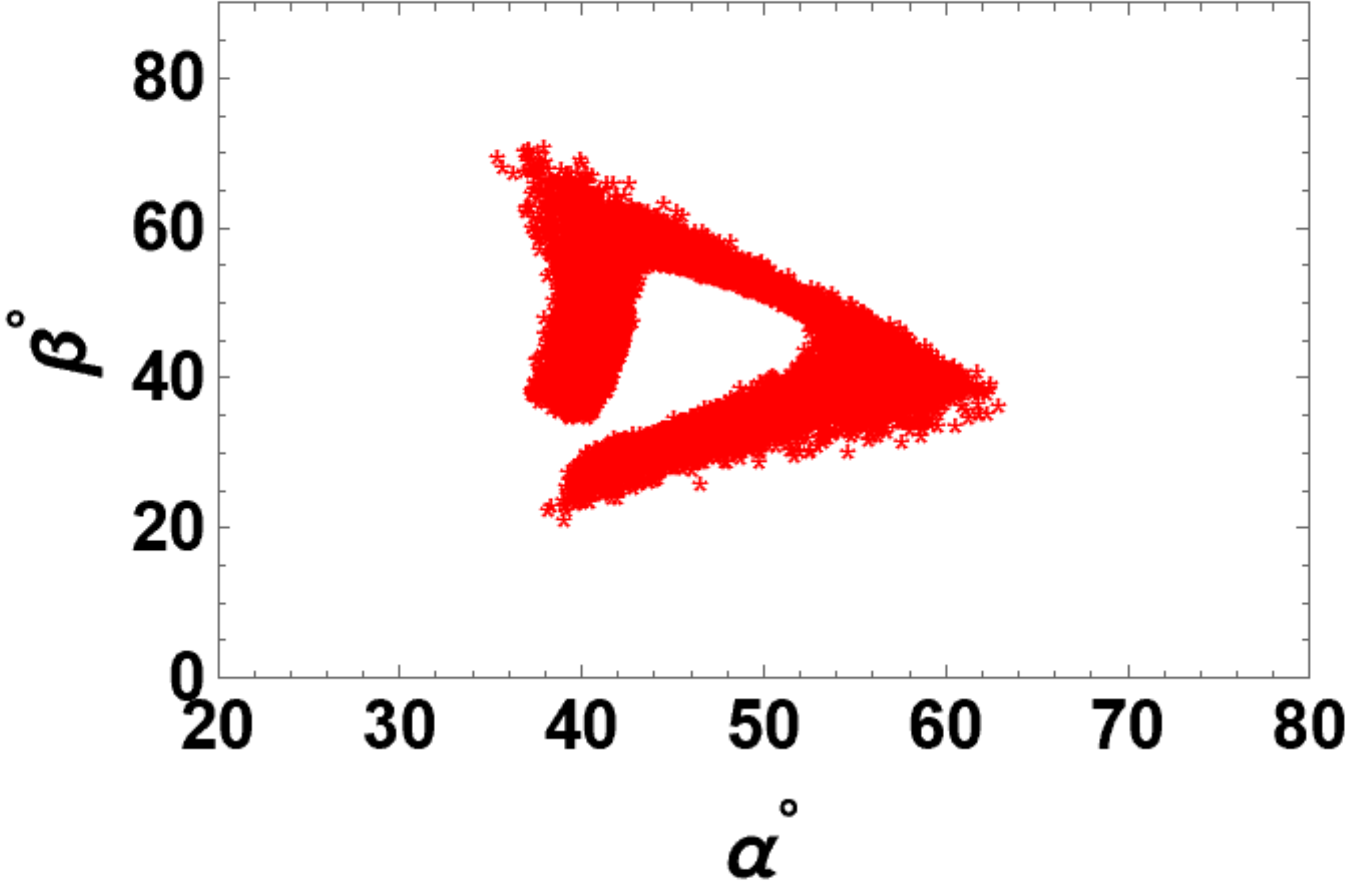}\label{fig:Type-A1 alpha vs beta.}} 
    \subfigure[]{\includegraphics[width=0.25\textwidth]{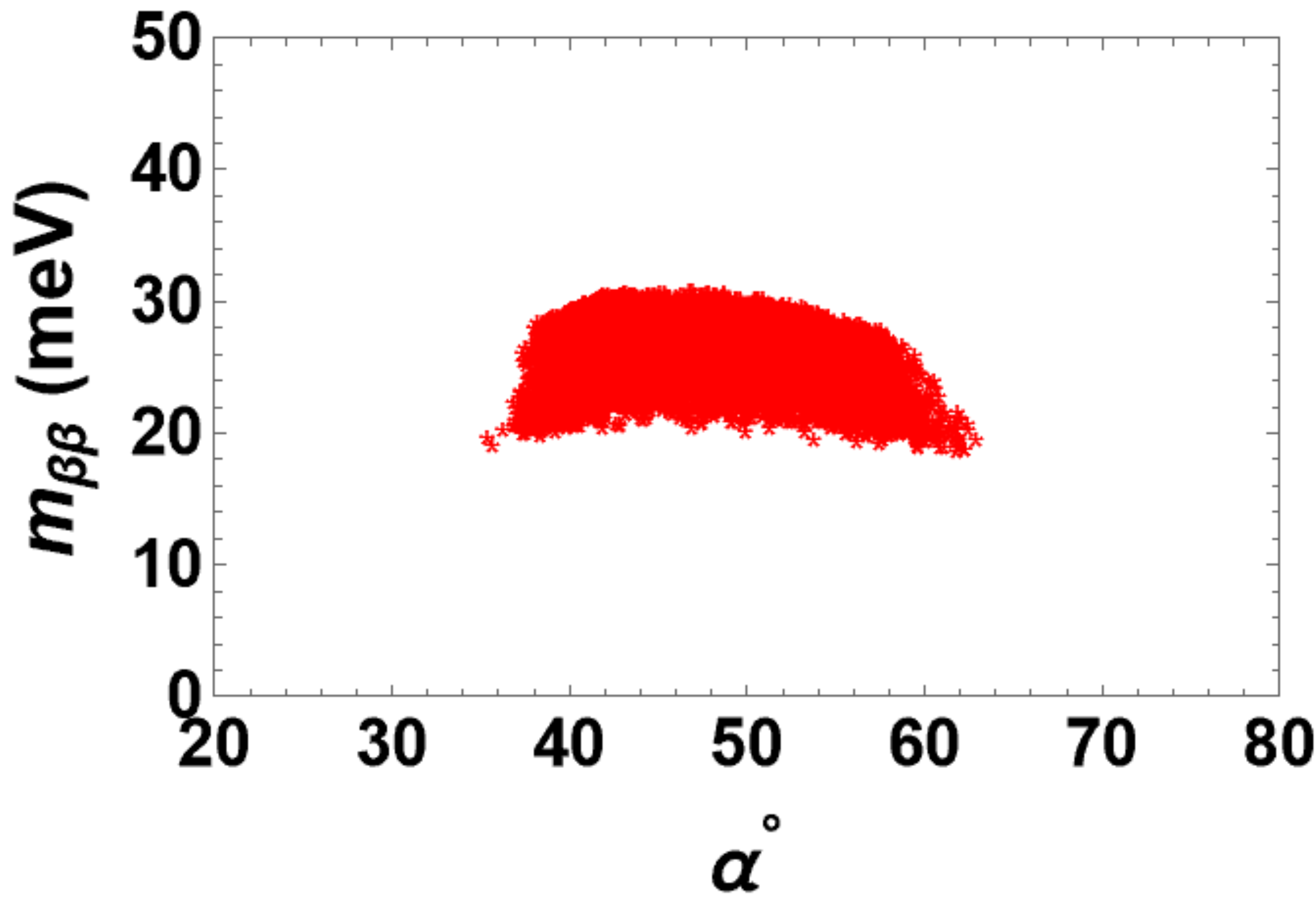}} 
\caption{Correlation plots between (a) $\alpha$ and $\beta$ (b) $\alpha$ and $m_{\beta\beta}$. For the graphical analysis of Type-A1 texture, we set $\theta_{23}\,=\,[39.6^{\circ}\,-\,51.8^{\circ}]$ and $\delta$ $[108^{\circ}\,-\,404^{\circ}]$.}
\label{fig:Type-A1}
\end{figure}

We shall highlight one texture from each family as an example. We start with the Type-\textbf{A} family which contains eleven textures. To illustrate the formalism, we pluck the Type- \textbf{A1} as an example. This texture shelters the following two correlations,

\begin{eqnarray}
\text{Re}\,[m_{ee}]&=&-\text{Im}\,[m_{e\mu}] \label{first relation for A1},\\
\text{Re}\,[m_{e\mu}]&=&-\text{Im}\,[m_{e\tau}],\label{second relation for A1}
\end{eqnarray}

which are two transcendental equations, and we solve these two equations using Newton-Raphson method to predict $\alpha$ and $\beta$. For this, we set initial values $\alpha_0$ and $\beta_0$ at $50^{\circ}$ and $60^{\circ}$ respectively. In addition, we fix $m_3$ and $\delta$ at $0.06\,eV$ and $270^{\circ}$ respectively, whereas $\Delta\,m_{ij}^2$ and $\theta_{ij}$'s are fixed at their best-fit values\,~\cite{n}. This results in $\alpha =47.86^{\circ} $ and $\beta=52.44^{\circ}$. The details of the numerical results, for Type-A1 along with the other ten textures from the same family are elaborated in the Table\,\ref{table:Values of Type-A textures}. To see the predictions for $\alpha$ and $\beta$ in a more comprehensive way for Type-\textbf{A1}, we adopt a graphical analysis, for which, instead of choosing specific numerical values, we generate sufficient amount of random numbers for an specific input parameter which are consistent within the $3\sigma$ bound as per Ref\,~\cite{n}. We ensure that

\begin{eqnarray}
&6.82\times 10^{-5}\, eV^2 \leq\Delta\,m_{21}^2\leq \,8.03\times 10^{-5}\,eV^2,&\\
&2.428\times 10^{-3}\,eV^2\leq \Delta\,m_{31}^2\leq\, 2.597\times 10^{-3}\,eV^2,&\\
&31.31^{\circ}\leq \theta_{12}\leq 35.74^{\circ},&\\
&8.19^{\circ}\leq \theta_{13}\leq\,8.89^{\circ},&\\
&39.6^{\circ}\leq \theta_{23}\leq\,51.9^{\circ}.&\\
&108^{\circ}\leq \delta\leq\,404^{\circ}.&
\end{eqnarray}

\begin{figure*}[!]
  \centering
    \subfigure[]{\includegraphics[width=0.24\textwidth]{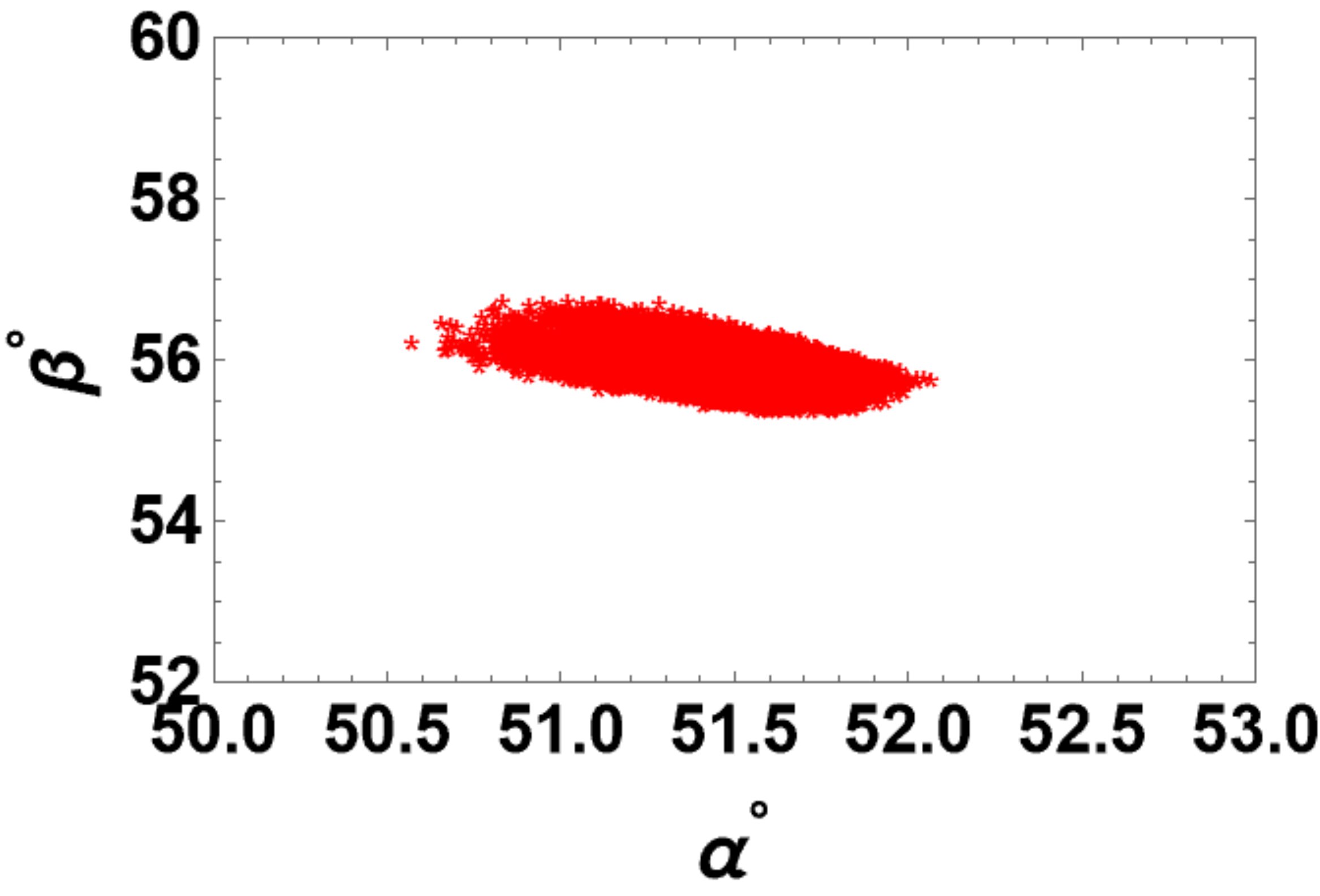}} 
    \subfigure[]{\includegraphics[width=0.24\textwidth]{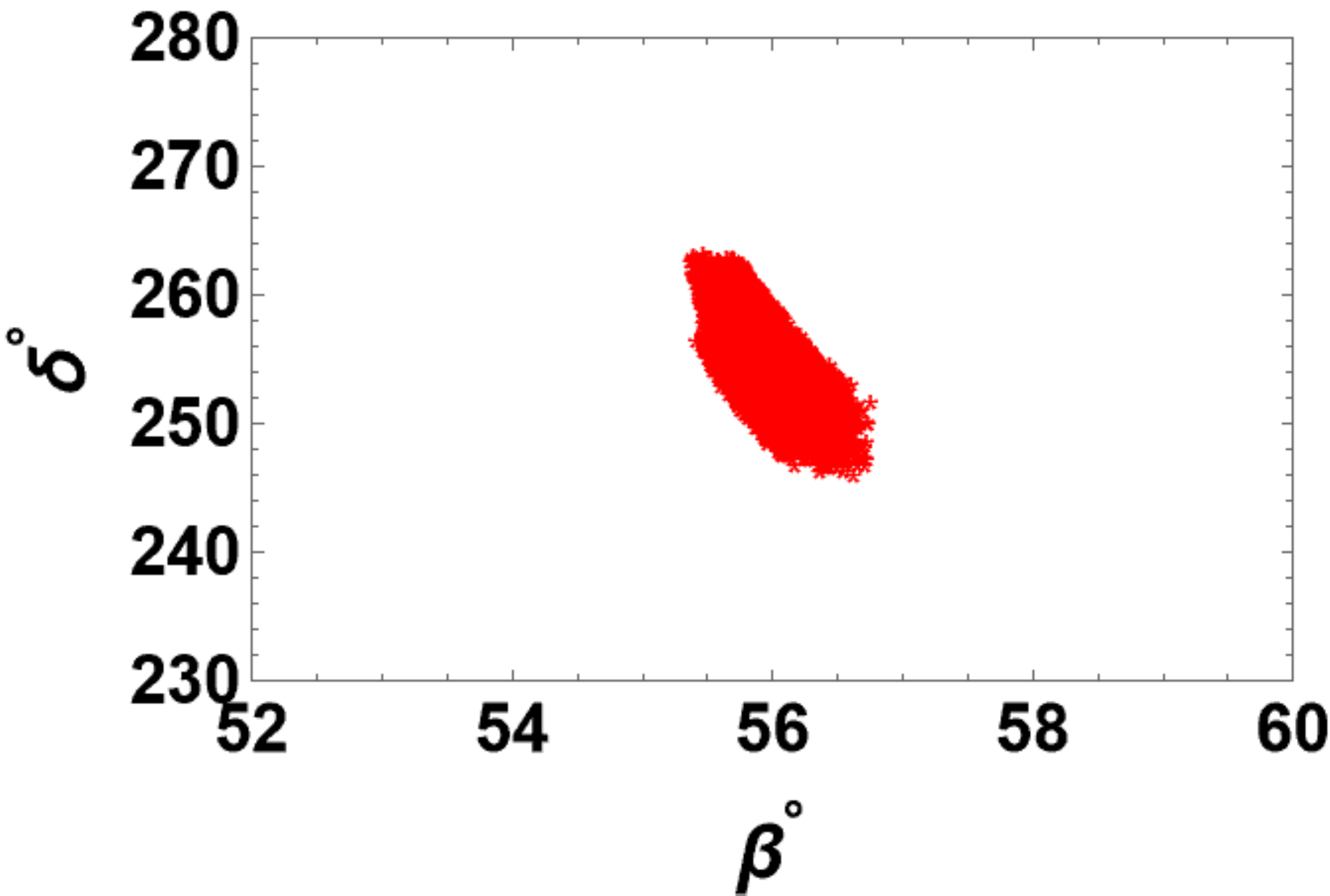}} 
    \subfigure[]{\includegraphics[width=0.24\textwidth]{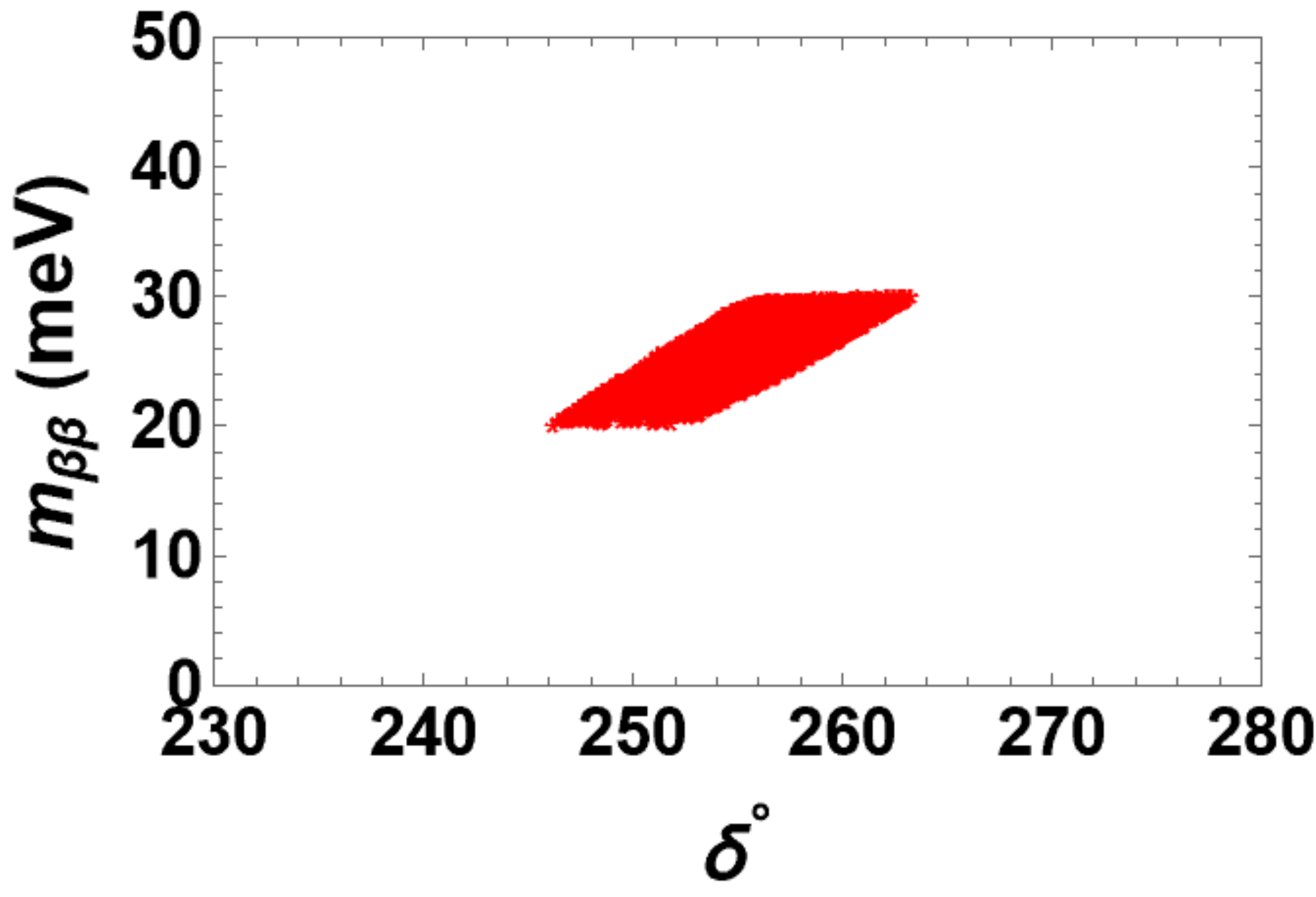}}
    \caption{Correlation plots between (a) $\alpha$ and $\beta$ (b) $\beta$ and $\delta$ (c) $\delta$ and $m_{\beta\beta}$. For the graphical analysis of Type-B1 texture, we set $\theta_{23}\,=\,[39.6^{\circ}\,-\,51.8^{\circ}]$.}
\label{fig:Type-B1}
\end{figure*}

\begin{figure*}[!]
  \centering
    \subfigure[]{\includegraphics[width=0.24\textwidth]{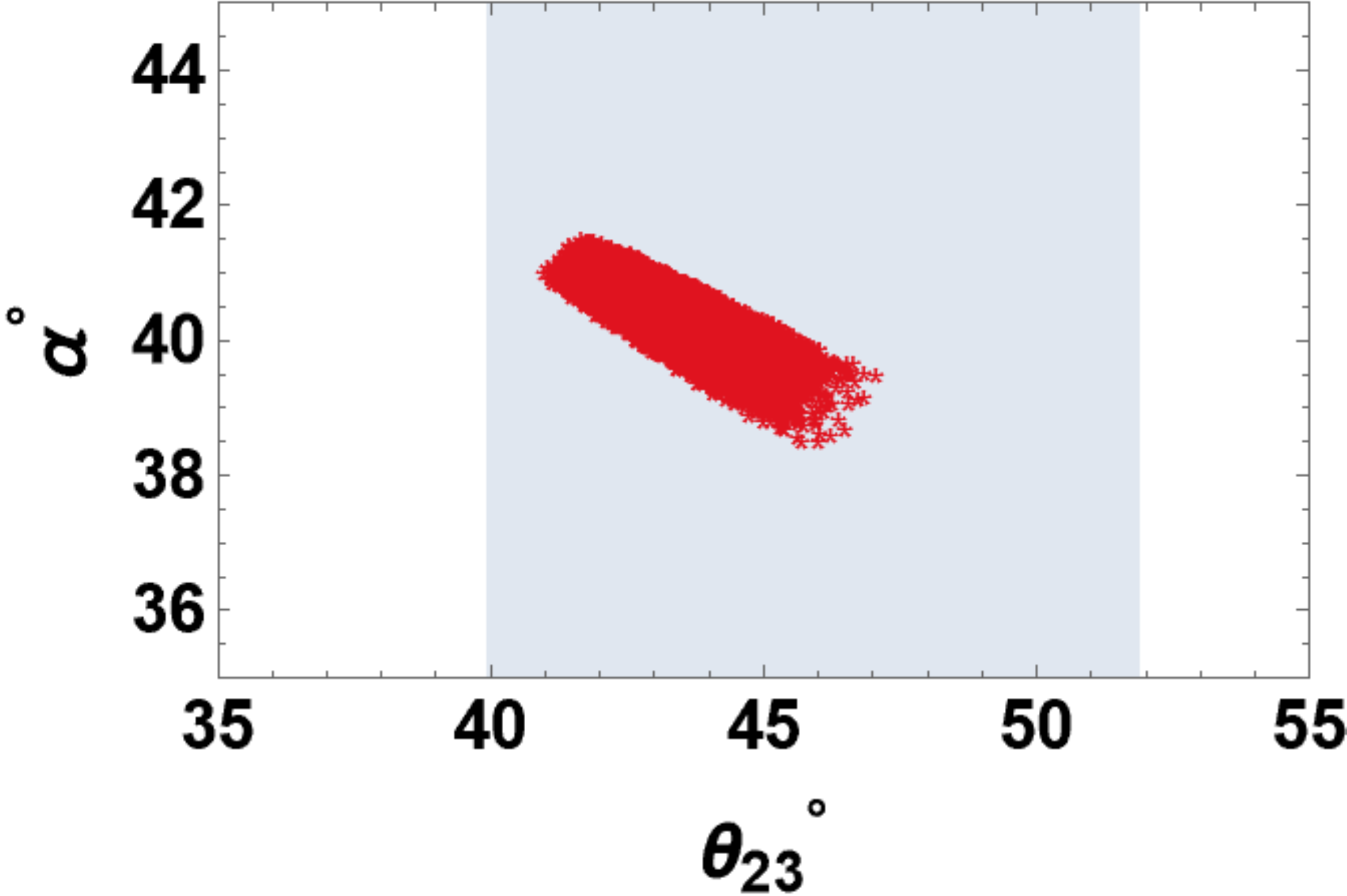}} 
    \subfigure[]{\includegraphics[width=0.24\textwidth]{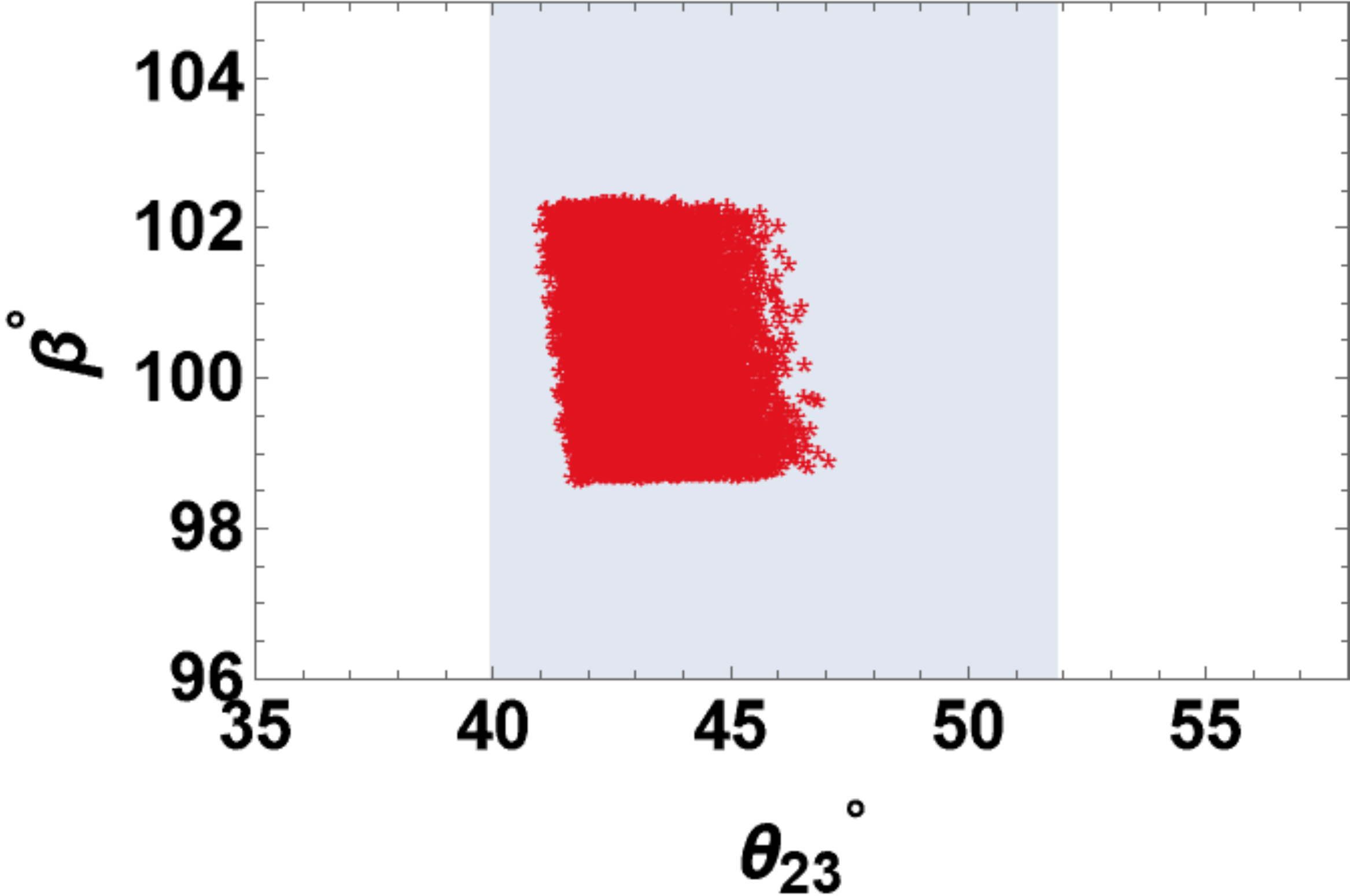}} 
    \subfigure[]{\includegraphics[width=0.24\textwidth]{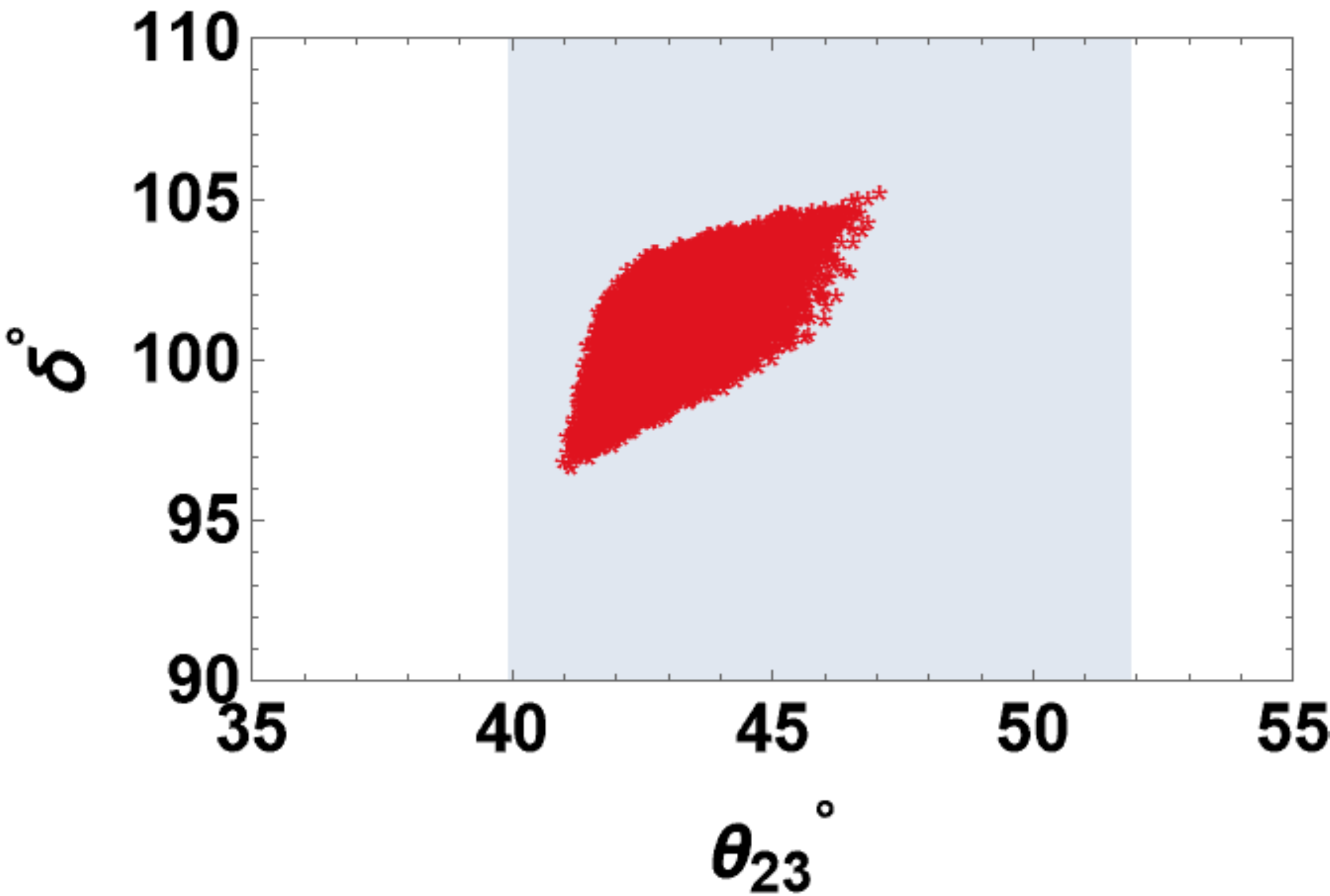}}
    \subfigure[]{\includegraphics[width=0.24\textwidth]{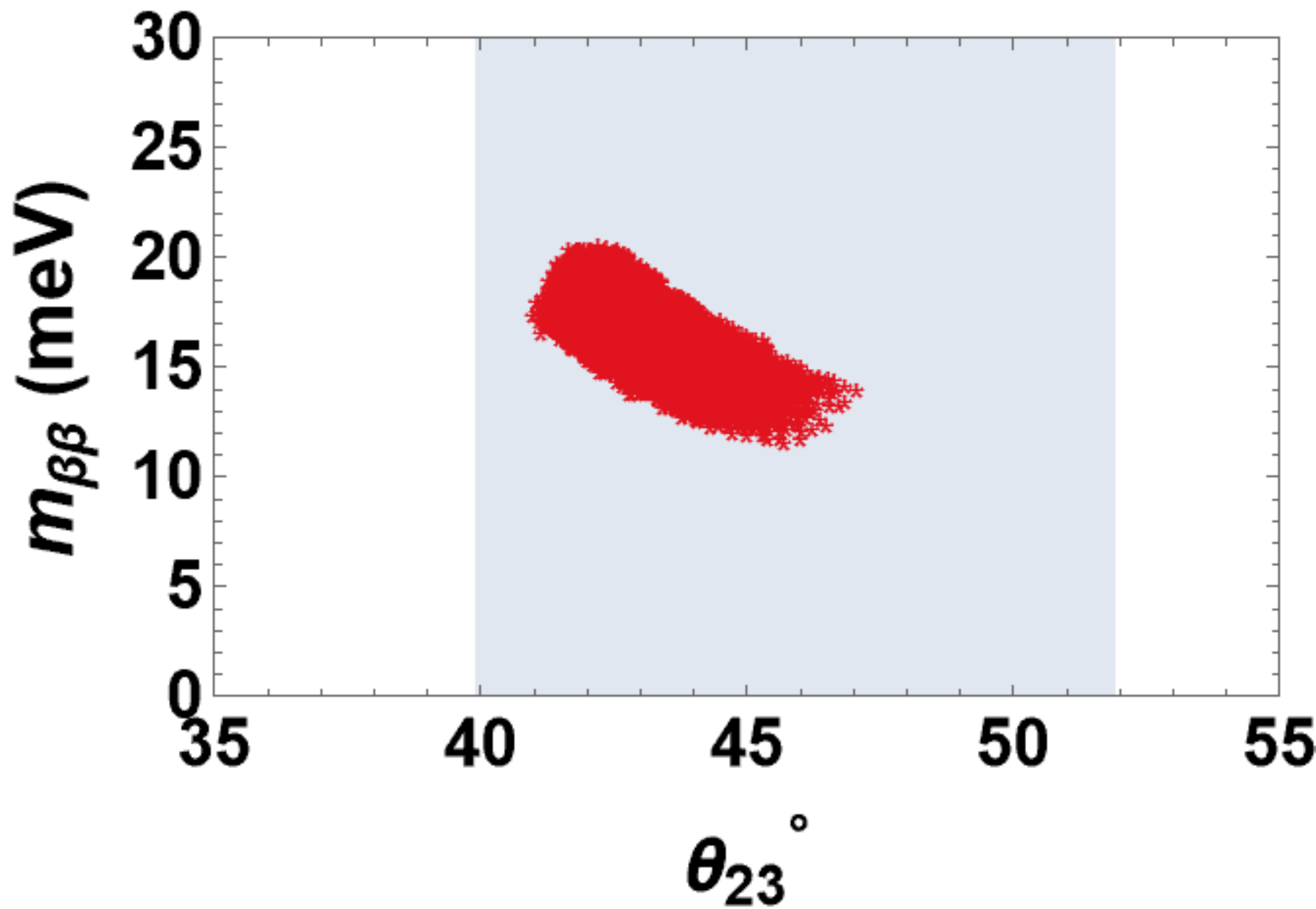}}
    \caption{Correlation plots between (a) $\theta_{23}$ and $\alpha$ (b) $\theta_{23}$ and $\beta$ (c) $\theta_{23}$ and $\delta$ (d) $\theta_{23}$ and $m_{\beta\beta}$. The blue strip represent the $3\sigma$ range of the observational  parameter $\theta_{23}$.}
\label{fig:Type-C1}
\end{figure*}

\begin{figure}[!]
  \centering
    \subfigure[]{\includegraphics[width=0.25\textwidth]{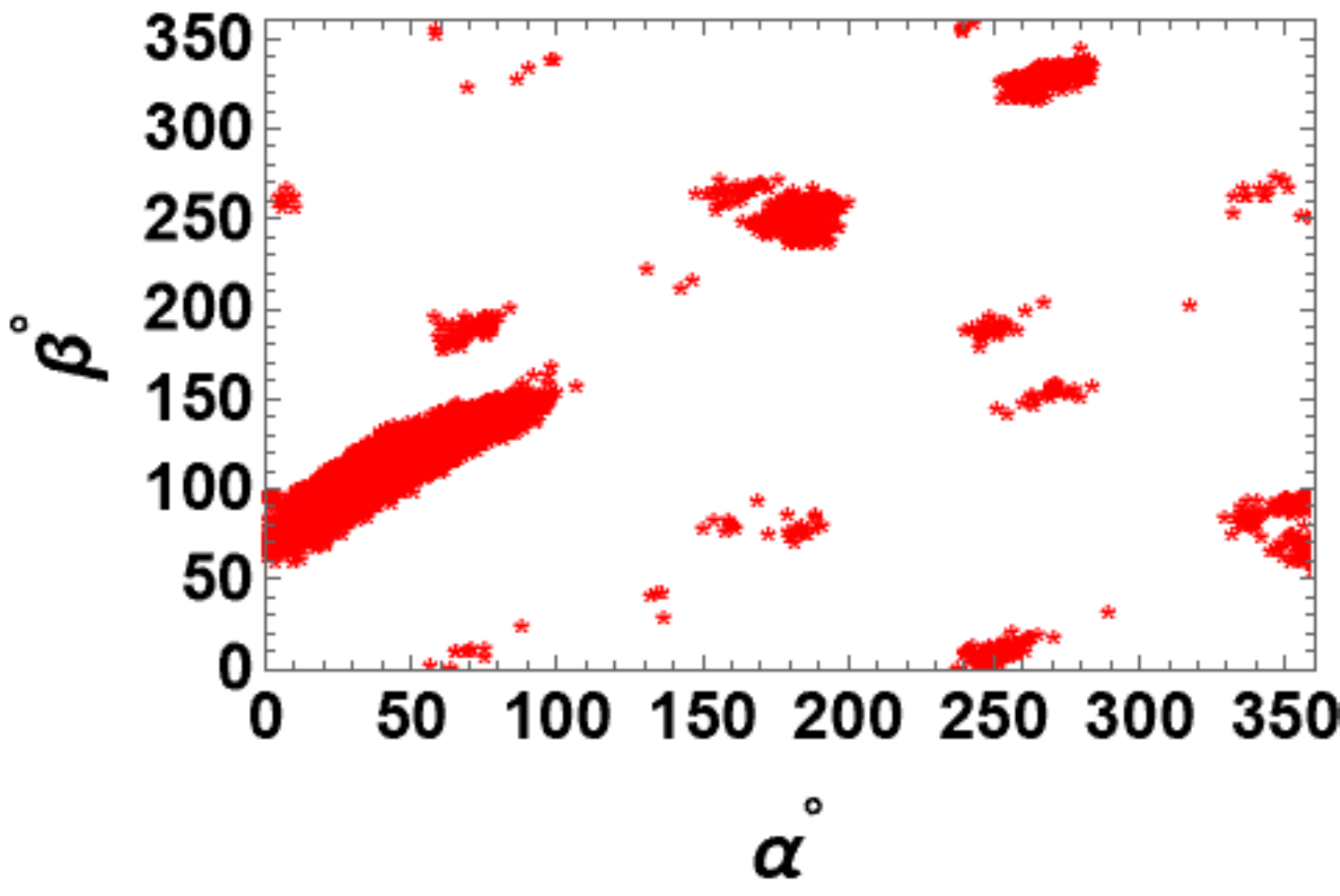}\label{fig:Type-A1 alpha vs beta.}} 
    \subfigure[]{\includegraphics[width=0.25\textwidth]{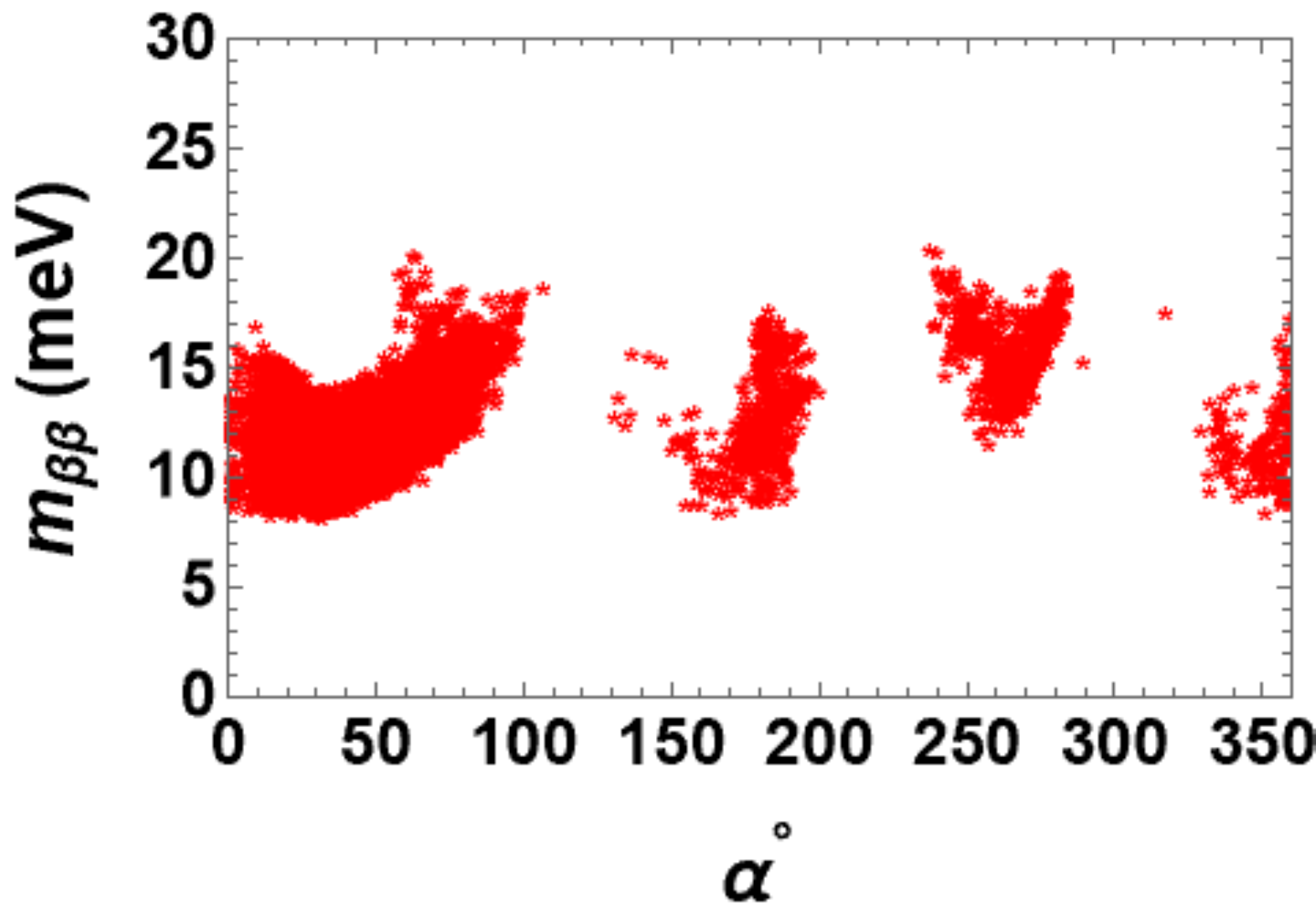}} 
\caption{Correlation plots between (a) $\alpha$ and $\beta$ (b) $\alpha$ and $m_{\beta\beta}$. For the graphical analysis of Type-A1 texture, we set $\theta_{23}\,=\,[39.6^{\circ}\,-\,51.8^{\circ}]$ and $\delta$ $[108^{\circ}\,-\,404^{\circ}]$.}
\label{fig:Type-P1}
\end{figure}

\begin{figure*}[!]
  \centering
    \subfigure[]{\includegraphics[width=0.24\textwidth]{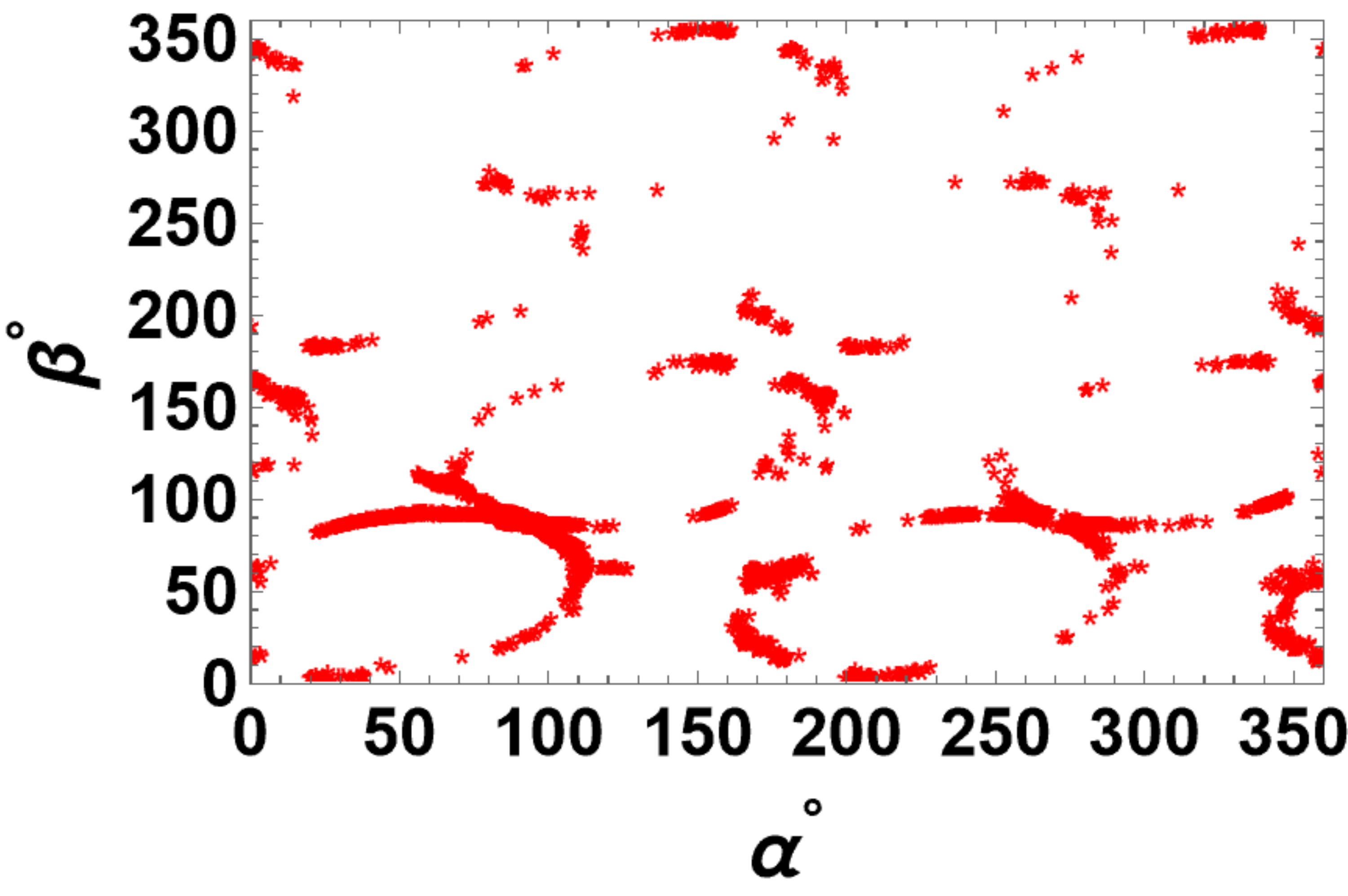}} 
    \subfigure[]{\includegraphics[width=0.24\textwidth]{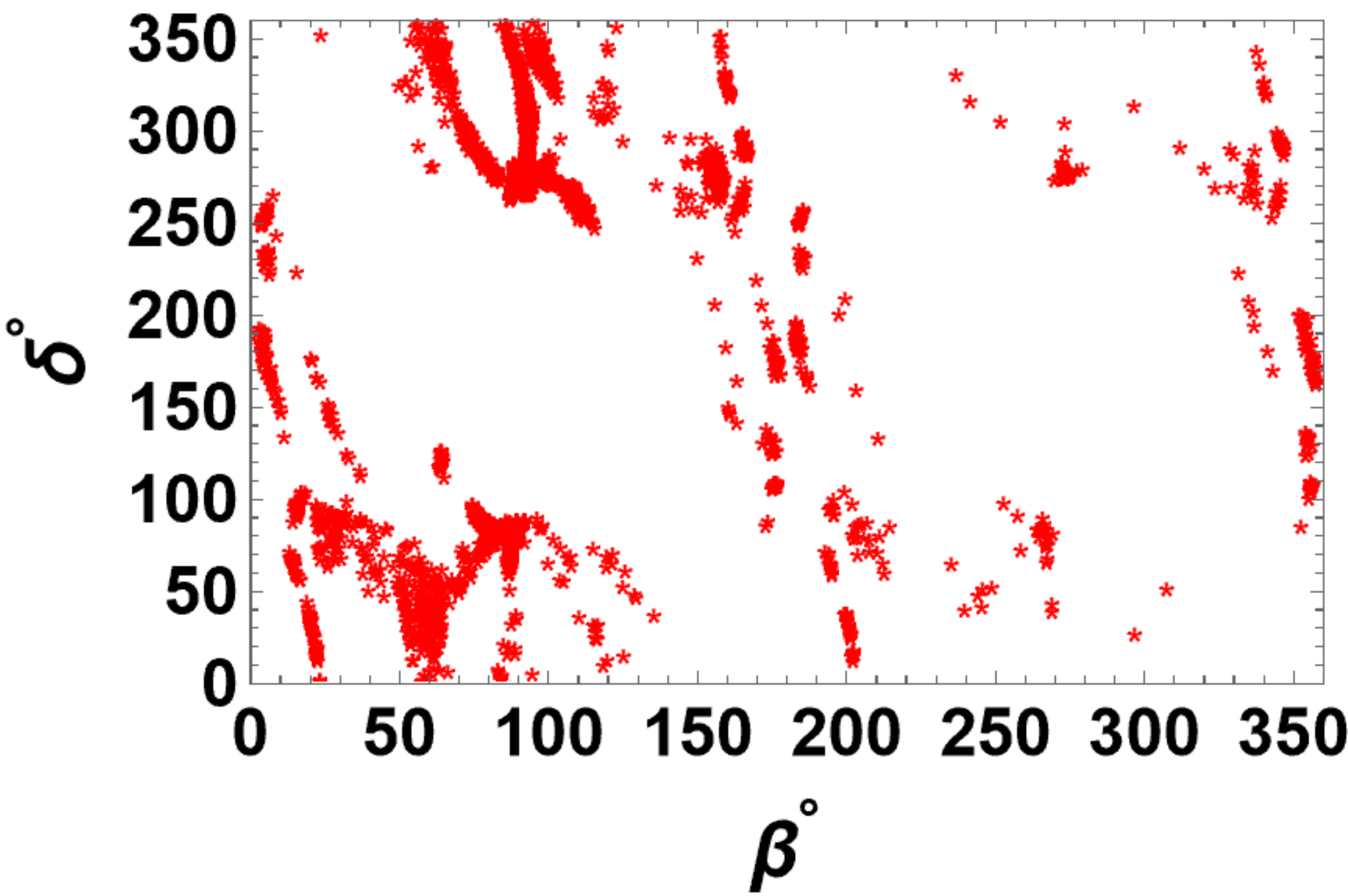}} 
    \subfigure[]{\includegraphics[width=0.24\textwidth]{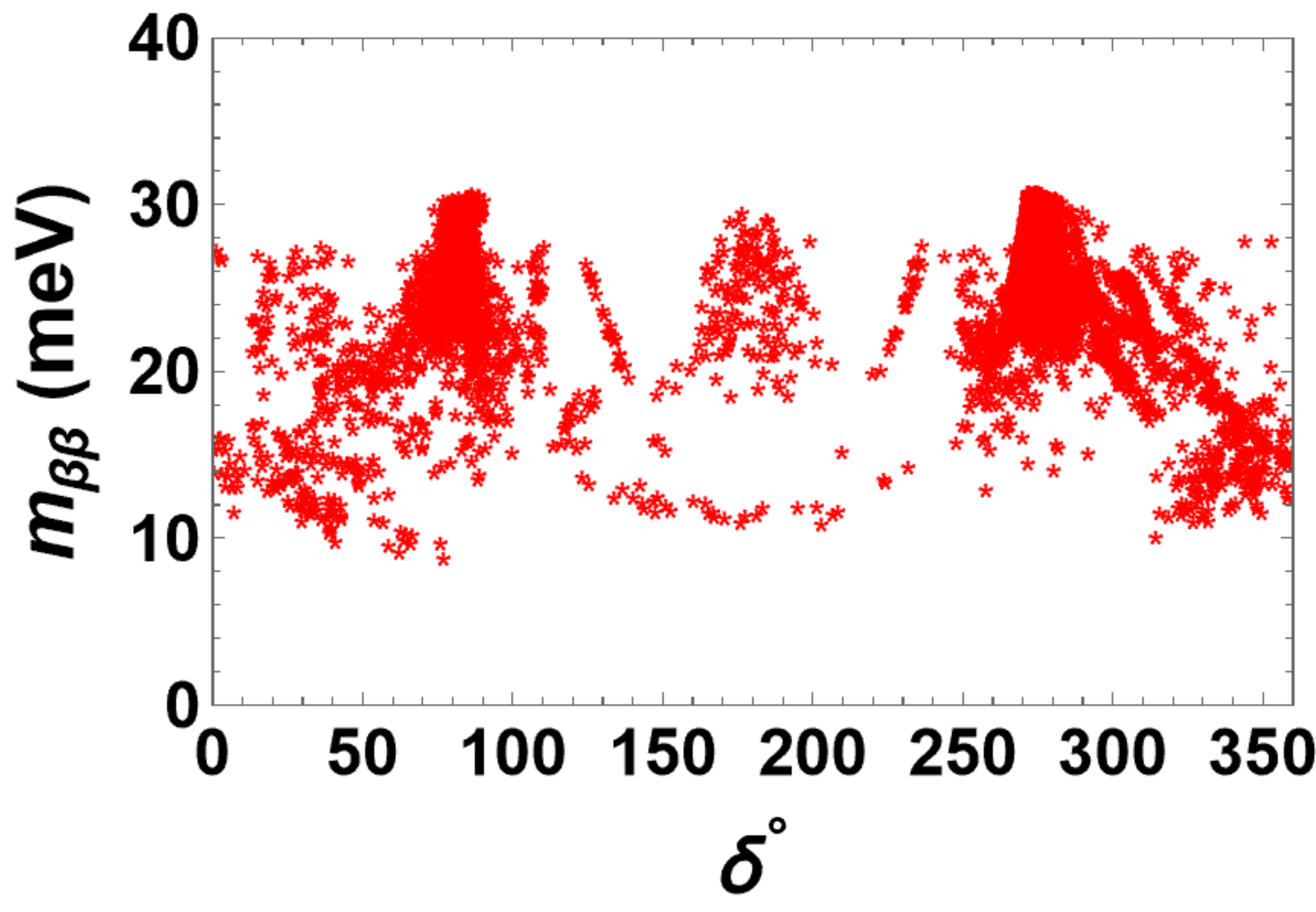}}
    \caption{Correlation plots between (a) $\alpha$ and $\beta$ (b) $\beta$ and $\delta$ (c) $\delta$ and $m_{\beta\beta}$. For the graphical analysis of Type-B1 texture, we set $\theta_{23}\,=\,[39.6^{\circ}\,-\,51.8^{\circ}]$.}
\label{fig:Type-Q1}
\end{figure*}

\begin{figure*}[!]
  \centering
    \subfigure[]{\includegraphics[width=0.24\textwidth]{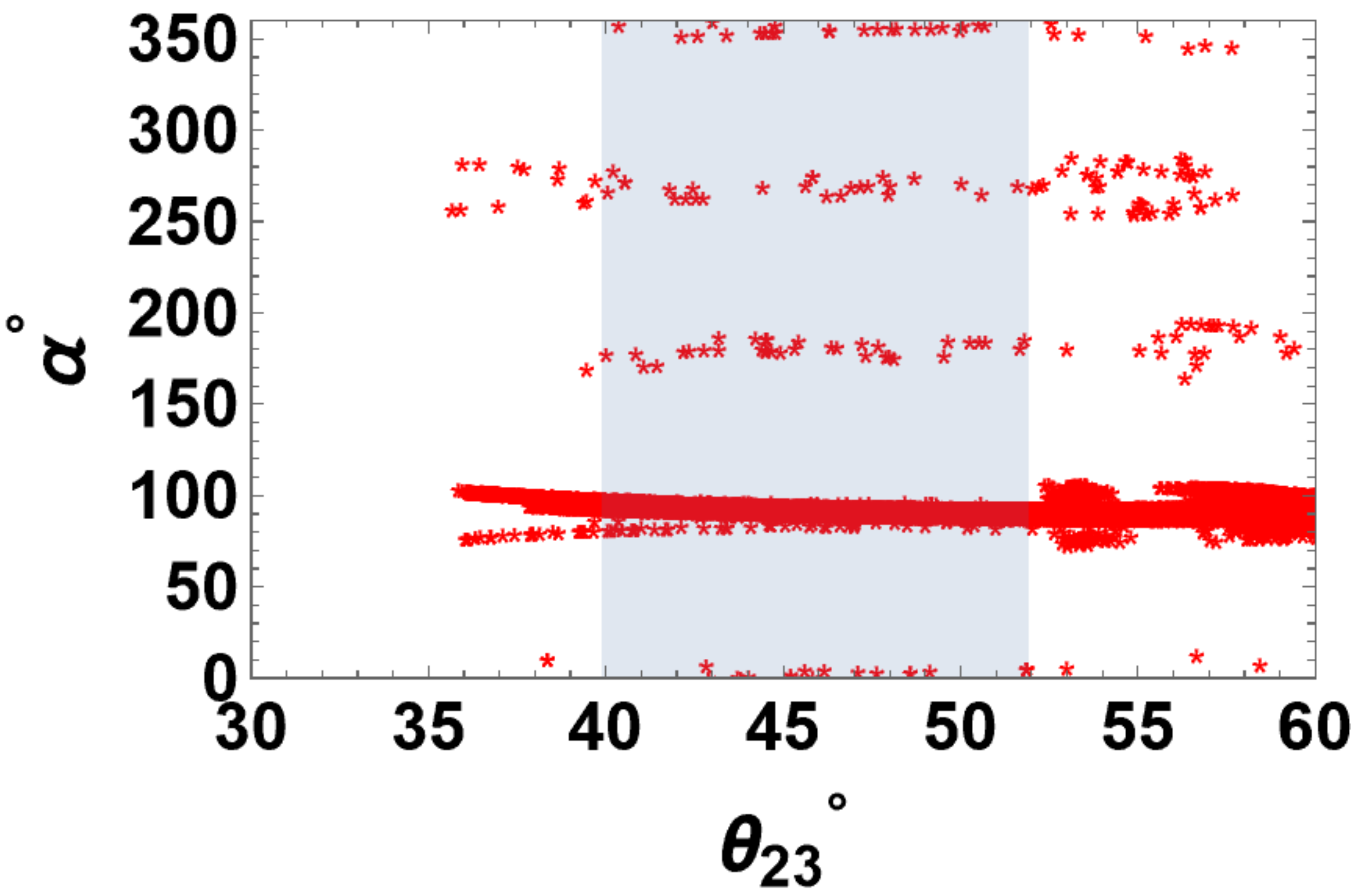}} 
    \subfigure[]{\includegraphics[width=0.24\textwidth]{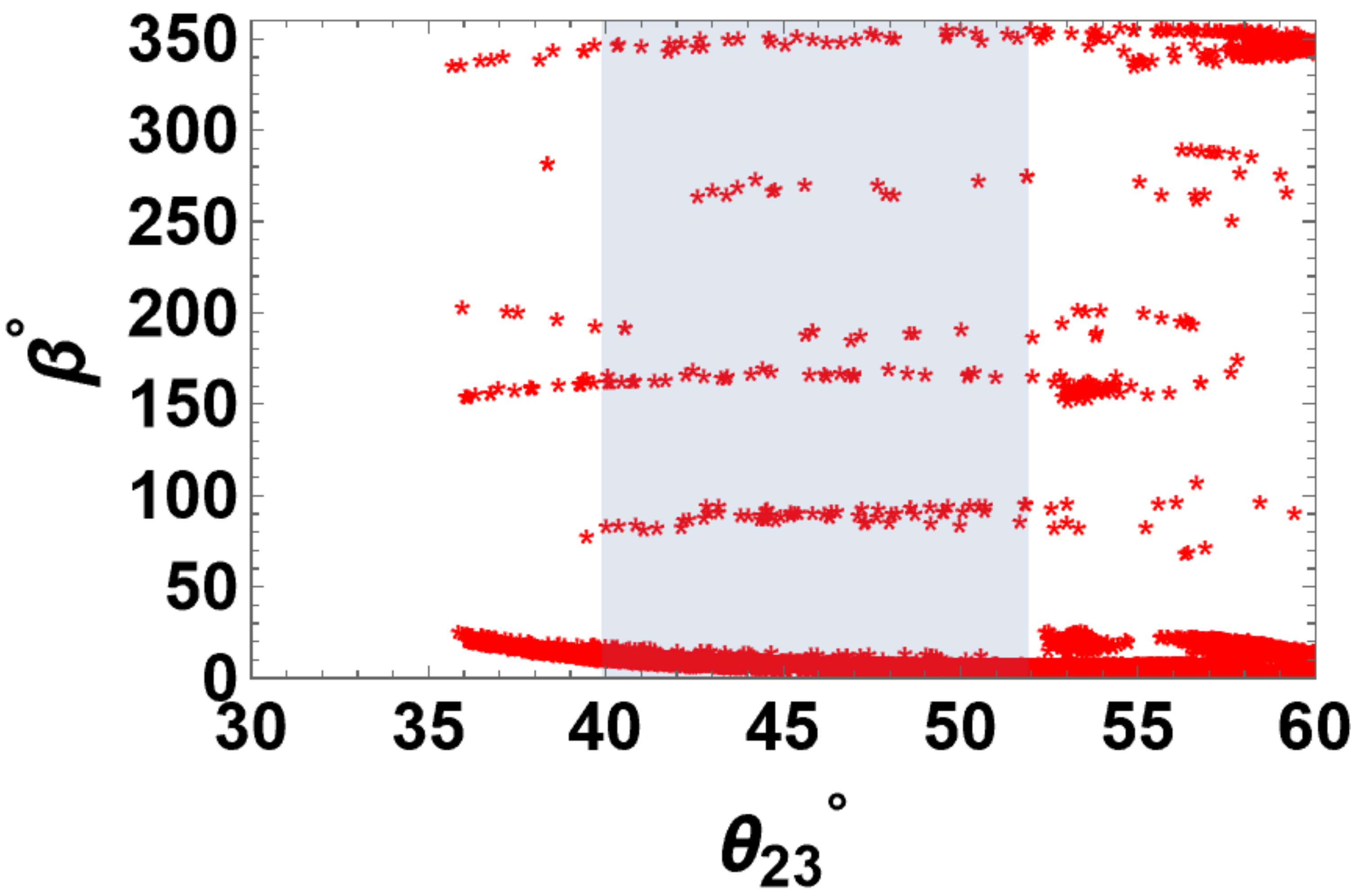}} 
    \subfigure[]{\includegraphics[width=0.24\textwidth]{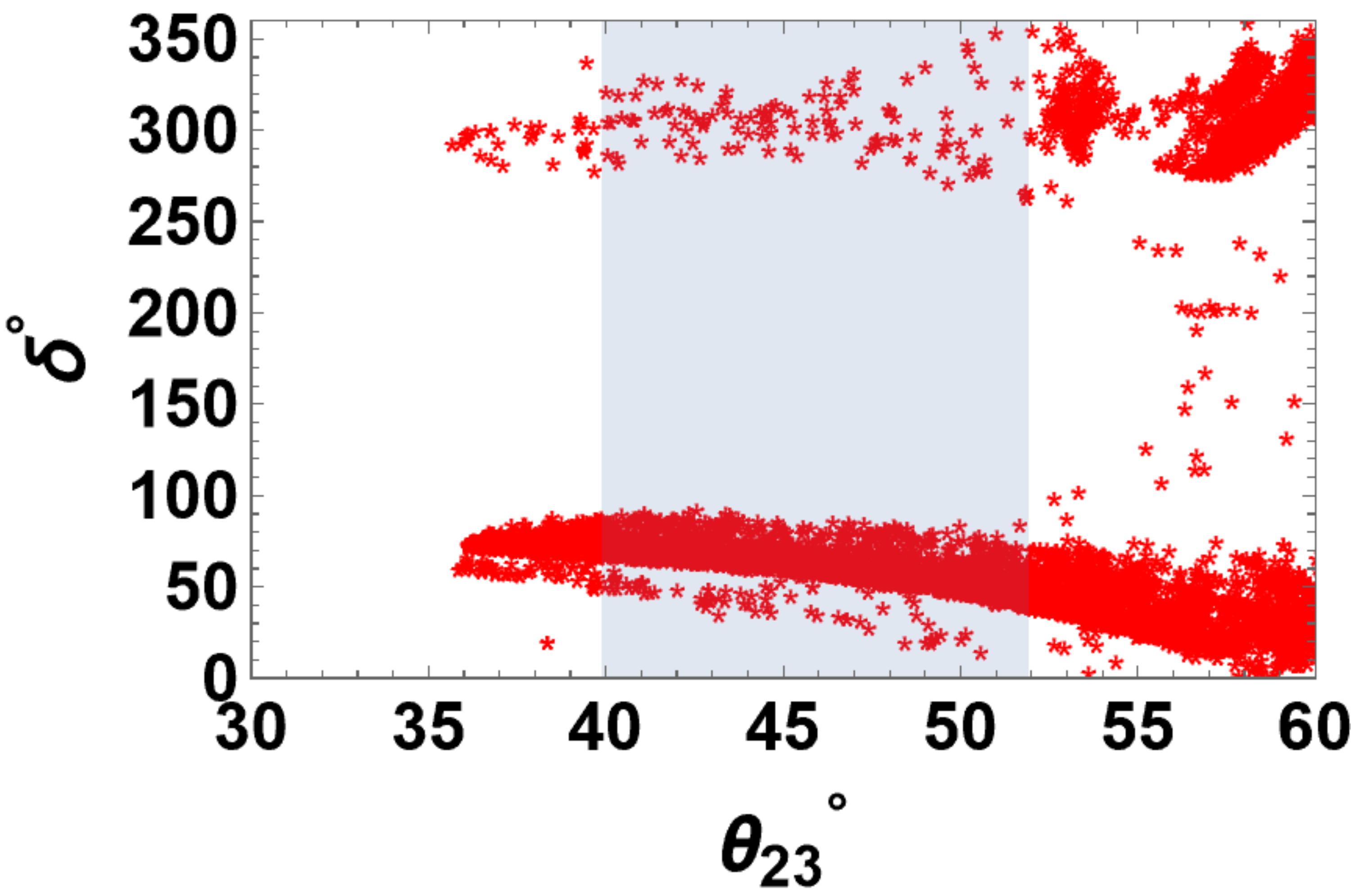}}
    \subfigure[]{\includegraphics[width=0.24\textwidth]{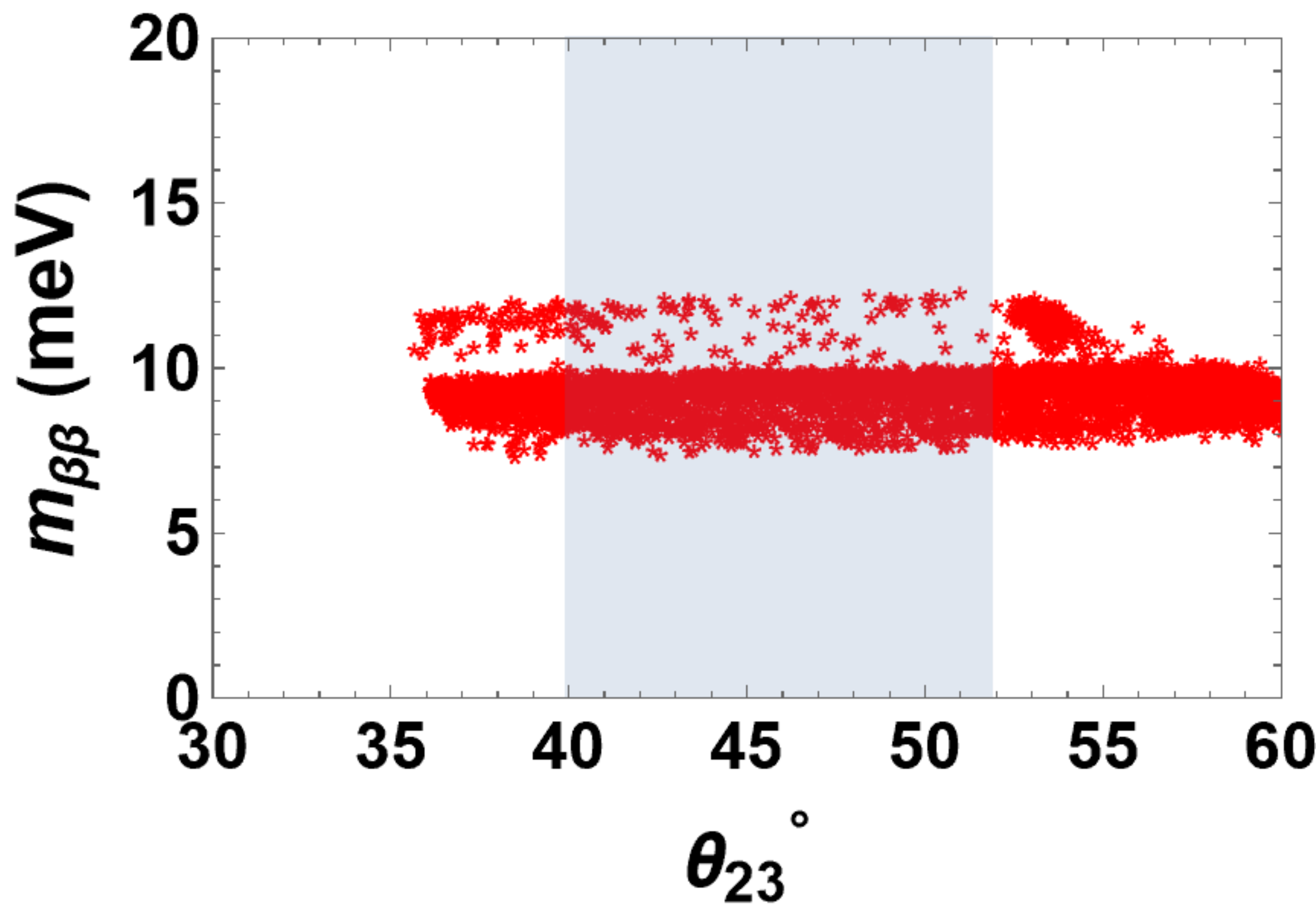}}
    \caption{Correlation plots between (a) $\theta_{23}$ and $\alpha$ (b) $\theta_{23}$ and $\beta$ (c) $\theta_{23}$ and $\delta$ (d) $\theta_{23}$ and $m_{\beta\beta}$. The blue strip represent the $3\sigma$ range of the observational  parameter $\theta_{23}$.}
\label{fig:Type-R1}
\end{figure*}

We choose the other input parameter, $0.055 \,eV\leq m_3\leq 0.06\,\,eV$ and this assures, $(m_1\,+\,m_2\,+\,m_3\leq0.12\,eV)$\,\cite{Planck:2018vyg}\footnote{We wish to state that for the graphical analysis shown in Figs.\,(\ref{fig:Type-A1})-(\ref{fig:Type-R1}), we strictly adhere to the input ranges prescribed here. However, for Type-B1 and Type-Q1, the input $\delta$ is not required and similarly, for Type-C1 and Type-R1, both $\delta$ and $\theta_{23}$ are redundant.}. Following this, we solve Eqs\,(\ref{first relation for A1}) and (\ref{second relation for A1}) and generate sufficient data points as predictions for $\alpha$ and $\beta$ which are illustrated graphically in Fig. \ref{fig:Type-A1}). We see that the predictions of the two Majorana phases are constrained.

Similarly in Type-\textbf{B} category, we generate twelve textures. For example, Type-\textbf{B1} texture underlines the following relations,

\begin{eqnarray}
\text{Im}\,[m_{ee}]&=&-3 \,\text{Re}\,[m_{ee}]\label{B1 first equation},\\
\text{Re}\,[m_{e\mu}]&=&-\text{Im}\,[m_{e\mu}],\\
2\,\text{Re}\,[m_{e\tau}]&=&-\text{Im}\,[m_{e\tau}]\label{B1 third equation}. 
 \end{eqnarray}

We shall exploit these three transcendental equations to predict $\delta$ in addition to $\alpha$ and $\beta$. With respect to a certain choice of the initial values $\alpha_0$, $\beta_0$ and $\delta_0$, the above three relations predict three physical parameters $\alpha$, $\beta$ and $\delta$ as $51.93^\circ$, $55.51^\circ$ and $259.99^\circ$ respectively. The detail of the required initial values of the said parameters along with  the numerical results related to predictions for all textures under Type-B family are elaborated in Table\,\ref{table:Values of Type-B textures}. As before, to analyse graphically the credibility of Type-B1 texture, we solve the above three Eqs.\,(\ref{B1 first equation})-(\ref{B1 third equation}) to generate sufficient random numbers for the parameters and obtain the correlation plots. From the plots, we identify the constrained bounds for the parameters $\alpha$, $\beta$, $\delta$ and $m_{\beta\beta}$\,(see Fig.\,\ref{fig:Type-B1}).

The Type-\textbf{C} family carries five textures. Out of which,  the Type-\textbf{C1} texture is exemplified. It deals with the following four relations, 
 
\begin{eqnarray}
3\,\text{Im}\,[m_{\mu\mu}]&=& \text{Im}\,[m_{e\tau}],\\
\text{Re}\,[m_{e\mu}]&=&-\text{Re}\,[m_{e\tau}],\\
\text{Im}\,[m_{e\tau}]&=&-\text{Re}\,[m_{ee}],\\
\text{Im}\,[m_{\tau\tau}]&=& 2 \,\text{Im}\,[m_{\mu\tau}],
\end{eqnarray}

and hence, can predict four observables: $\alpha$, $\beta$, $\delta$ and $\theta_{23}$. We take proper choices of the initial values and Type-C1 predicts: $\alpha$ as $41.41^\circ$ , $\beta$ as $100.3^\circ$, $\delta$ as $99.81^\circ$ and $\theta_{23}$ as $41.16^\circ$. The details of the input and predicted values of the Type-C textures can be found in Table \ref{table:Values of Type-C textures}. Following the same approach as before, we obtain the correlation plots and identify the constrained regions for the parameters $\alpha$, $\beta$, $\delta$, $\theta_{23}$ and $m_{\beta\beta}$ (see Fig.\,\ref{fig:Type-C1}).

Similarly, we find twelve textures in Type-\textbf{P} family. To exemplify, we see that the Type-\textbf{P1} texture deals with the following relations,

\begin{eqnarray}
2\, |m_{ee}| &=& |m_{\mu\mu}|\label{Type-P Eq1},\\
|m_{e\mu}|&=&|m_{e\tau}|\label{Type-P Eq2},
\end{eqnarray}

and predict the Majorana phases $\alpha$ and $\beta$ as $6.77^{\circ}$ and $80.77^{\circ}$ respectively, with respect to certain initial choices $\alpha_0$ and $\beta_0$. The details of the numerical inputs and predicted values of the parameters under Type-P textures are presented in the Table \ref{table:Values of Type-P textures}. To explore the possible regions of $\alpha$, $\beta$ and $m_{\beta\beta}$ for which Type-P1 texture is valid, we generate the correlation plot\,(see Fig.\,\ref{fig:Type-P1}).

We organize seven textures in Type-\textbf{Q} category. We put forward Type-\textbf{Q1} as an example which shelters three relations as shown,

\begin{eqnarray}
3\,|m_{ee}|&=& \text{arg}\,[m_{e\mu}],\\ 
2 \,|m_{e\mu}|&=& \text{arg}\,[m_{\tau\tau}],\\ 
|m_{e\tau}|&=&2\, \text{arg}\,[m_{\mu\mu}],
\end{eqnarray} 

which are exploited to predict $\alpha$, $\beta$ and $\delta$. With certain choices of initial values, we estimate $\alpha$, $\beta$ and $\delta$ as $92.57^{\circ}$, $26.88^{\circ}$ and $165.01^{\circ}$ respectively. The associated numerical inputs and results for all seven textures under Type-Q family are presented in Table\,\ref{table:Values of Type-Q textures}. To visualise graphically, we generate the correlation plots and identify the constrained regions for the above said parameters (see Fig.\,\ref{fig:Type-Q1}).

In fine, we work out six textures in Type-\textbf{R} category. For example, in Type-\textbf{R1} texture, the following transcendental equations 

\begin{eqnarray}
3\,|m_{ee}|&=&\text{arg}\,[m_{\tau\tau}],\\
|m_{e\mu}|&=&2\,\text{arg}\,[m_{\tau\tau}],\\ 
|m_{e\tau}|&=&2\,\text{arg}[m_{ee}],\\
|m_{\mu\tau}|&=&3\,\text{arg}\,[m_{ee}],
\end{eqnarray}

coexist and hence the texture may predict four observable parameters: $\alpha$, $\beta$, $\delta$ and $\theta_{23}$ which are predicted as $98.64^{\circ}$, $18.69^{\circ}$, $40.59^{\circ}$ and $40.78^{\circ}$ respectively. These predictions are subjected to certain initial choices of the said parameters and are described in Table \ref{table:Values of Type-R textures}. The latter highlights the initial values as well as predictions related to all the textures of Type-Q family. We generate the corresponding correlation plots plots and identify the constrained bounds for the above mentioned parameters\,(see Fig.\,\ref{fig:Type-R1}).

Here, we emphasize that the predictions regarding the observational parameters with respect to relevant textures detailed  in Tables\,(\ref{table:Values of Type-A textures})-(\ref{table:Values of Type-R textures}) are sensitive to the initial values of the observational parameters. It is important to note that the numerical value associated with the prediction of a given texture is not unique; rather, there exist multiple viable solutions. For the Tables (\ref{table:Values of Type-A textures}) to (\ref{table:Values of Type-R textures}), we chose the inputs, namely $\Delta m_{21}^2$, $\Delta m_{31}^2$, $\theta_{12}$ and $\theta_{13}$ as the best-fit values\,~\cite{n} with an exception of the Type-Q5 texture in Table.\,(\ref{table:Values of Type-Q textures}), where $\Delta m_{31}^2$ is set to $2.53\times 10^{-3}\, eV^2$ (within $1\sigma$ range~\cite{n}). Furthermore, for all the above mentioned tables, we fix $m_3$ at  $0.06\,eV$. For Tables\,(\ref{table:Values of Type-A textures}) and (\ref{table:Values of Type-P textures}), the Dirac CP phase $\delta$ is fixed at $270^{\circ}$. Except for Tables \,(\ref{table:Values of Type-C textures}) and (\ref{table:Values of Type-R textures}), the mixing angle $\theta_{23}$ is fixed at its best-fit value~\cite{n}. Following the predictions obtained from Tables\,(\ref{table:Values of Type-A textures})-(\ref{table:Values of Type-R textures}), we predict the numerical values of the parameter $m_{\beta\beta}$ corresponding to fifty-three proposed textures in Tables\,(\ref{table:Values of DBB under GP}) and (\ref{table:Values of DBB under EP}).

The Majorana phases are undetermined in the experiments. In principle, they can take any value between $0^\circ$ to $360^\circ$. However, in the Ref\cite{deGouvea:2008nm}, it is shown that the Majorana CP phases can be chosen to lie within $[0^{\circ}-180^{\circ}]$ whereas the Dirac CP phase $\delta$ can take any value within $[0^{\circ}-360^{\circ}]$. We see that the prediction for $\beta$ from Type-Q5 texture contradicts with Ref\cite{deGouvea:2008nm}. In the present work, we attempt to constrain the values of the Majorana phases so that in future experiments they can be tested. We highlight that the neutrino mass matrix textures obtained in the present work are consistent with the data obtained from oscillation experiments. All the predicted values of $\delta$ and $\theta_{23}$ under Type-B,C,Q,R textures lie within the $3\sigma$ range of neutrino oscillation data. However, the Majorana phases are unobservable in the oscillation experiments. The viability of the proposed textures is subjected to the future experiments as discussed in section\,\ref{Introduction}.

\section{Symmetry Realization \label{Symmetry Realization}}

All the neutrino mass matrix textures proposed in the present work are obtained from phenomenological ground by following the bottom up approach. Though these textures are model independent, it is expected that they can be derived from the first principle. To make the discussion more fruitful, we try to realize the underlying possible symmetry groups working behind one specific texture from each category. 

\subsection{Type-A3}
The Type-A3 texture deals with two constrained relations,
\begin{eqnarray}
\text{Re}\,[m_{\tau\tau}]&=&-2\,\text{Re}\,[m_{e\tau}] \label{equnA31},\\  
\text{Im}\,[m_{e\tau}] &=& \text{Im}\,[m_{\mu\tau}]\label{equnA32}.
\end{eqnarray}

To derive this particular texture, we extend the SM by introducing two additional $SU(2)_L$ doublet, three $SU(2)_L$ triplet Higgs fields and three $SU(2)_L$ singlet fields. The right-handed neutrino field transforms as triplet under $A_4$. We consider that the right-handed neutrino field and the additional $SU(2)_L$ triplet Higgs fields contribute to the effective neutrino mass matrix via Type-I and Type-II mechanisms respectively. The additional symmetry $Z_2$ is incorporated along with $A_4$ in order to restrict the unwanted terms that are allowed by the latter. The transformation of all the field contents of our proposed model with their respective assignments under the group $SU(2)_L \times A_4 \times Z_2$ are summarised in Table\,\ref{Field Chart of Type-A3}.

The $SU(2)_L \times A_4 \times Z_2$ invariant Yukawa Lagrangian can be constructed in the following manner,

\begin{eqnarray}
- \mathcal{L}_Y &=& y_e (\bar{D}_{l_L}\Phi)_1\,e_{R_1} + y_{\mu} (\bar{D}_{l_L}\Phi)_{1'} \,\mu_{R_{1''}}+y_{\tau} (\bar{D}_{l_L}\Phi)_{1''}\, \tau_{R_{1'}} +\nonumber\\&& y_{D} \{(\bar{D}_{l_L}\tilde{\Phi})_{3_S}\, \nu_{R}+(\bar{D}_{l_L}\tilde{\Phi})_{3_A} \,\nu_{R}\}+\frac{1}{2}\,y_{R_1} (\bar{\nu}_{R}\,\nu^c_R)_1 \,\eta \,+\nonumber\\&& \frac{1}{2}\,y_{R_2} (\bar{\nu}_{R}\,\nu^c_R)_3\,\kappa+\frac{1}{2}\,y_{R_3} (\bar{\nu}_{R}\,\nu^c_R)_3\,\xi\,+\, y_{T} (\bar{D}_{l_L}\, D_{l_L}^c)_3\,\Delta\,\nonumber\\&&+\,h.c.,
\label{Yukawa Lagrangian}
\end{eqnarray}

where, $\tilde{\Phi}=i \tau_2 \Phi^{*}$. The products rules under $A_4$ group can be found in \ref{appendix b}. It is important to mention that we choose a specific VEV alignment $(\Phi_i)$ as $\langle\Phi\rangle_{0}\,=\,v\,(1,\omega, \omega^2)^T$ (a similar choice is mentioned in ref ~\cite{t}), where, $\omega=e^{2 \pi i/3}$ and $\omega^2=e^{- 2 \pi i/3}$. In the symmetry basis, such a choice of the VEVs lead us to the charged lepton mass matrix of the following form,

\begin{equation}
M_L = v \begin{bmatrix}
y_e & y_\mu & y_\tau\\
y_e\omega & y_\mu & y_\tau \omega^2\\
y_e \omega^2 & y_\mu & y_\tau \omega\\
\end{bmatrix}.
\end{equation}

On the other hand, the Dirac neutrino mass matrix is obtained as shown below,

\begin{equation}
M_D = 2\,y_{D} \,v \begin{bmatrix}
0 & 0 & \omega\\
\omega^2 & 0 & 0\\
0 & 1 & 0\\
\end{bmatrix}.
\end{equation}

We assume that the fields $\kappa$ and $\xi$ attain the VEV alignments  $v_{\kappa}(0,0,1)^T$ and $v_{\xi}(1,0,0)^T$ while $\langle\eta\rangle= v_\eta$ such that the right-handed neutrino mass matrix takes the following form,

   \begin{equation}
M_R = \begin{bmatrix}
p & q & 0\\
q & p & r\\
0 & r & p\\
\end{bmatrix},
\end{equation}

where, $p = y_{R_1} v_\eta$, $q= y_{R_2} v_\kappa$, $r=y_{R_3} v_\xi$. For the sake of simplicity, we take the following choices: $q= 2p$ and $r=3p$.

After the redefinition of some parameters, we obtain the neutrino mass matrix from Type-I seesaw mechanism in the following way,

\begin{eqnarray}
M_{T_1} &=& - M_D M^{-1}_R M^T_D\nonumber\\
        &=& \begin{bmatrix}
-a\omega^2 & 2a & -a\omega\\
2a & \frac{-8}{3}a\omega & \frac{-2}{3}a\omega^2\\
-a\omega & \frac{-2}{3}a\omega^2 & \frac{a}{3}\\
\end{bmatrix}
\end{eqnarray}

Assuming that the Higgs triplet takes VEVs along the direction $\langle\Delta\rangle_0 = v_\Delta(0,1,1)^T$, the Type-II contribution to the effective neutrino mass matrix takes the following form,

 \begin{equation}
M_{T_2} = d \begin{bmatrix}
0 & 1 & 1\\
1 & 0 & 0\\
1 & 0 & 0\\                  
\end{bmatrix}, \,\,\,\,\,\,\,\,\,\,\,\,\,\,\,\,\,\,\,\,\,\, \text{where,}\,\,\,\,\,d=y_T v_\Delta
\end{equation}

After taking the contributions from Type-I and Type-II seesaw, the effective neutrino mass matrix can be written as,

\begin{eqnarray}
M_\nu &= M_{T_1}+M_{T_2} 
&=\begin{bmatrix}
-a\omega^2 & 2a+d & -aw+d\\
2a+d & \frac{-8}{3}a\omega & \frac{-2}{3}a \omega^2\\
-a\omega +d & \frac{-2}{3}a\omega^2 & \frac{a}{3}\\                  
\end{bmatrix}.
\end{eqnarray}

To move from symmetry basis to flavour basis, we need to diagonalise $M_l$ in the following way,
\begin{equation}
M^{diag}_l = U_l^{\dagger} M_l U_r,
\end{equation}

where, \begin{equation}
U_l = \frac{1}{\sqrt{3}} \begin{bmatrix}
1 & 1 & 1\\
1 & \omega^2 & \omega\\
1 & \omega & \omega^2\\                  
\end{bmatrix} \,\,\,\text{and} \,\,\,U_r=\begin{bmatrix}
0 & 0 & 1\\
1 & 0 & 0\\
0 & 1 & 0\\                  
\end{bmatrix}.
\end{equation}

In the flavour basis, the effective neutrino mass matrix is also perturbed by the charged lepton correction ($U^{\dagger}_{l_L} M_{\nu_s}  U^*_{l_L}$) which is expressed in the following manner,

\begin{equation}
M^f_\nu =
\begin{bmatrix}
A+iB &  C+iD  &  \mathbf{E}+i\mathbf{F} \\
C+iD & G+iH  &  J+i\mathbf{F} \\
E+iF & J+iF & -\mathbf{2E}-2iF \\
\end{bmatrix}.
\label{Type-A3 mass matrix}
\end{equation}

From the neutrino mass matrix appearing in Eq.\,\ref{Type-A3 mass matrix}, we see that the mass matrix elements $(M^f_{\nu})_{e\tau}$, $(M^f_{\nu})_{\mu\tau}$ and $(M^f_{\nu})_{\tau\tau}$ exhibit the relations prescribed in Eqs.\,(\ref{equnA31}) and (\ref{equnA32}). Thus, the above mass matrix, $M_{\nu}^f$ resembles Type-A3 texture.

\subsection{Type-B8}

The Type-B8 texture shelters three constrained relations: 

\begin{eqnarray}
\text{Re}\,[m_{e\mu}]&=&\text{Im}\,[m_{e\tau}],\\
\text{Re}\,[m_{ee}]&=&-\text{Im}\,[m_{\mu\tau}],\\
\text{Im}\,[m_{\mu\mu}]&=&- \text{Im}\,[m_{\tau\tau}].
\end{eqnarray}

 To derive the Type-B8 texture, we extend the field content of the SM by introducing right-handed neutrinos\,$(\nu_{e_R},\nu_{\mu_R},\nu_{\tau_R})$, two scalar singlets\,($\kappa$ and $\varsigma$) and a scalar triplets\,($\Delta$). We summarise the transformation properties of the field content associated with $\Delta\,(27)$ symmetry in Table\,\ref{Field Chart of Type-B8}. We construct the $SU(2)_L \times \Delta\,(27)$ invariant Lagrangian in the following way,

\begin{eqnarray}
- \mathcal{L}_Y &=& y_e (\bar{D}_{l_L}H)_{1_{00}}\,e_{R_{1_{00}}} + y_{\mu} (\bar{D}_{l_L}H)_{{1_{20}}} \,\mu_{R_{1_{10}}}+\,y_{\tau}\,(\bar{D}_{l_L} H)_{{1_{10}}} \tau_{R_{1_{20}}} +\nonumber\\&& y_1\,(\bar{D}_{l_L}\tilde{H})_{{1_{00}}}\,\nu_{e_{R_{1_{00}}}} + y_2 (\bar{D}_{l_L}\,\tilde{H})_{1_{10}} \nu_{\mu_{R_{1_{20}}}} + y_3 (\bar{D}_{l_L}\tilde{H})_{1_{20}}\,\nu_{\tau_{R_{1_{10}}}}\,\nonumber\\&&+\frac{1}{2}\,y_{R_1}\,(\bar{\nu}_{\tau_R}\,\nu^c_{\tau_R})_{1_{20}}\,\kappa_{1_{10}} + \frac{1}{2}\,y_{R_2}\,[(\bar{\nu}_{e_{R}}\,\nu^c_{\mu_{R}})\,+ (\bar{\nu}_{\mu_{R}}\,\nu^c_{e_{R}})]_{1_{20}}\,{\kappa}_{1_{10}}\nonumber\\&& +\, y_{T_2}\,\bar{D}_{l_{L_{3}}}\,( D_{l_L}^c\,{\Delta})_{3^*_{s_{1}}}+\, h.c.
\label{Yukawa Lagrangian Type-B8}
\end{eqnarray}

The product rules under $\Delta\,(27)$ are given in \ref{appendix a}. We choose the complex vacuum alignment $\langle H \rangle_{\circ}=v_{H}\,(w,1,1)^{T}$ as per ref \cite{Branco:1983tn} and obtain the charged lepton mass matrix as shown below,

\begin{equation}
 M_{l} = v_{\phi}
 \begin{bmatrix}
 \omega^2 \, y_{e} & \omega^2 \,y_{\mu} & \omega^2 \, y_{\tau}\\
 \omega^2 y_{e} &  \,y_{\mu} & \omega \,y_{\tau}\\
 \omega y_{e} &  \,y_{\mu} & \omega^2 \,y_{\tau}\\
 \end{bmatrix},
 \label{chargrd lepton mass matrix}
\end{equation}
where, $\omega= e^{i\,2\pi/3}$, $\omega^2= \omega^*$.

We diagonalise $M_l$ as $M^{diag}_{l}=\,U^{\dagger}_{l_L}M_{l}U_{l_R}$, where, $M^{diag}_{l}=\,\sqrt{3}\,v_H\,diag\,(y_e,\,y_{\mu},\,y_{\tau})$. The $U_{l_L}$ and $U_{l_R}$ can  be expressed in the following way,
\begin{eqnarray}
 U_{l_L} &=& \frac{1}{\sqrt{3}}
 \begin{bmatrix}
 \omega^2\,e^{i\zeta} & e^{i\psi} & \omega\,e^{i\varphi}\\
 e^{i\zeta} & \omega^2\,e^{i\psi}  & \omega\,e^{i\varphi} \\
 e^{i\zeta} & e^{i\psi}  & e^{i\varphi} \\
 \end{bmatrix}, 
 \label{Mass matrix 6}\\
 U_{l_R} &=& diag( e^{i\zeta},\, \omega\,^{i\psi},\,\omega^2\, e^{i\varphi}).
\end{eqnarray}

It is to be noted that the choices of eigenvectors are not unique and we include the arbitrary phases ($\zeta,\psi,\varphi$) in $U_{l_L}$ and $U_{l_R}$. It has many advantages which are highlighted in ref \cite{Dey:2022qpu}. To experience the Type-B8 texture, we set $\zeta=\frac{\pi}{2}$, $\psi=\pi$ and $\varphi=\frac{\pi}{2}$. 

Under the choice of vacuum alignment $\langle H \rangle_{\circ}=v_{H}\,(w,1,1)^{T}$ and $\langle\kappa\rangle_{\circ}=v_{\kappa}$, the Dirac neutrino mass matrix ($M_D$) and right handed neutrino mass matrix ($M_R$) takes the following form,

\begin{eqnarray}
 M_{D} &=&
 \begin{bmatrix}
  x\,\omega & y\,\omega & z\,\omega\\
 x & \omega^2\,y & \omega\,z \\
 x & \omega\,y  & \omega^2\, z \\
 \end{bmatrix},
 \label{Dirac Mass Matric M1}\\
 M_{R}&=&
 \begin{bmatrix}
 0 & q & 0\\
 q & 0 & 0 \\
 0 & 0  & p \\
 \end{bmatrix},
 \label{M_R for M1}
\end{eqnarray}

where, $x=y_1\,v_H$, $y=y_2\,v_H$, $z=y_3\,v_H$, $p=Y_{R_1}\,v_\kappa$ and $q=Y_{R_2}\,v_\kappa$.

The Type-I seesaw contributes as $M_{T_{1}}=\,-\,M_{D}M^{-1}_{R}M^{T}_{D}$. The choices of vacuum alignments $\langle\Delta\rangle_{\circ}=v_{\Delta}(1,1,1)^{T}$ and $\langle\chi\rangle_{\circ}=v_{\chi}$ for the $\Delta$ Higgs and scalar field $\chi$ give rise to the Type-II contribution as shown below,

\begin{equation}
 M_{T_2} = 
 \begin{bmatrix}
 r & 0 & 0\\
0 & r & 0 \\
  0 &  0  & r \\
 \end{bmatrix}, 
 \label{Type-II for M1}
\end{equation}

where, $r= y_{T_2}\,v_\Delta$. In symmetry basis, we construct the neutrino mass matrix as $M_{\nu_s}=M_{T_{1}}+ M_{T_{2}}$. In flavour basis, after taking the contribution from $U_{l_L}$, the neutrino mass matrix takes the following form,

\begin{equation}
M^f_\nu =
\begin{bmatrix}
\textbf{A}+i B &  \textbf{C}+iD  &  F+i\textbf{C} \\
C+iD & G+i\textbf{H}  &  J-i\textbf{A} \\
F+iC & J-iA &  -G-i\textbf{H} \\
\end{bmatrix}.
\end{equation}

\subsection{Type-C2}

The Type-C2 texture shelters three constrained relations: 

\begin{eqnarray}
\text{Re}\,[m_{ee}]&=&\text{Re}\,[m_{\mu\tau}]\\
\text{Re}\,[m_{e \mu}]&=&-\text{Re}\,[m_{e\tau}]\\
\text{Im}\,[m_{\mu \mu}]&=&\text{Im}\,[m_{\tau\tau}]\\
\text{Im}\,[m_{\mu \mu}]&=&- \text{Im}\,[m_{\mu\tau}]
\end{eqnarray}

 We try to derive the Type-C2 texture from the same Lagrangian appearing in Eq. \ref{Yukawa Lagrangian Type-B8} with a modification in the Type-II contribution part. In this regard we add a scalar singlet $\chi_{1_{10}}$ in the field content of the SM.  For the Type-C2 texture, we modify the Type-II part of the Lagrangian as shown below,

\begin{eqnarray}
- \mathcal{L}_{\text{Type-II}} &=& \frac{ y_{T_2}}{\Lambda}\,\bar{D}_{l_{L_{3}}}\,( D_{l_L}^c\,{\Delta})_{3^*_{s_{1}}}\chi_{1_{10}}+\, h.c.
\label{Yukawa Lagrangian Type-C2}
\end{eqnarray}

 The choices of vacuum alignments $\langle\Delta\rangle_{\circ}=v_{\Delta}(1,1,1)^{T}$ and $\langle\chi\rangle_{\circ}=v_{\chi}$ for the $\Delta$ Higgs and scalar field $\chi$ give rise to the Type-II contribution as shown below,

\begin{equation}
 M_{T_2} = 
 \begin{bmatrix}
 r & 0 & 0\\
0 & \omega r & 0 \\
  0 &  0  & \omega^2 r \\
 \end{bmatrix}, 
 \label{Type-II for C2}
\end{equation}

where, $r=\frac{T_2}{\Lambda}\,v_\Delta v_\chi$. In symmetry basis, the neutrino mass matrix is constructed as as $M_{\nu_s}=M_{T_{1}}+ M_{T_{2}}$. In flavour basis, we take the contribution from $U_{l_L}$, and the neutrino mass matrix takes the following form,

\begin{equation}
M^f_\nu =
\begin{bmatrix}
\textbf{A}+iB &  \textbf{C}+iD  &  -\textbf{C}+iE \\
C+iD & G+i\textbf{H}  &  \textbf{A}-i\textbf{H} \\
-\textbf{C}+iE & A-i\textbf{H} & J+i\textbf{H} \\
\end{bmatrix}.
\end{equation}

We highlight that the arbitrary phases appearing in $U_{l_L}$ and $U_{l_R}$ are set as $\zeta=\pi$, $\psi=2\pi$ and $\varphi=\frac{\pi}{2}$ to experience Type-C2 texture.

\subsection{Type-P4}

Under EP, the Type-P4 texture deals with two constrained relations:

\begin{eqnarray}
|m_{e\mu}|&=&|m_{e\tau}|\\
\text{arg}\,[m_{e\mu}]&=&-\text{arg}[m_{e\tau}]
\end{eqnarray}

 We try to experience this particular texture from $T_7$ symmetry group. We extend the SM by adding two scalar triplet and one anti-triplet. The transformation properties of the field content are summarised in Table \ref{Field Chart of Type-P4}. The product rules under $T_7$ group can be found in \ref{appendix c}. We construct the $SU(2)_L \times T_7$ invariant Yukawa Lagrangian in the following way,

\begin{eqnarray}
- \mathcal{L}_Y &=& y_e (\bar{D}_{l_L}\Phi)\,e_{R} + y_{\mu} (\bar{D}_{l_L}\Phi) \,\mu_{R}+y_{\tau} (\bar{D}_{l_L}\Phi)\, \tau_{R} + y_{D}(\bar{D}_{l_L}\tilde{\Phi})\,\nu_{R}\nonumber\\&&+\frac{1}{2}\,y_{R_1} (\bar{\nu}_{R}\,\nu^c_R)\chi\,+ \frac{1}{2}\,y_{R_2} (\bar{\nu}_{R}\,\nu^c_R)\,\kappa+\, y_{T_2} (\bar{D}_{l_L}\,\bar{D}_{l_L}^c)\,\Delta\,+\,h.c.,
\label{Yukawa Lagrangian p4}
\end{eqnarray}

where, $\tilde{\Phi}=i \tau_2 \Phi^{*}$. 

The specific choice of the vacuum expectation value $\langle\Phi\rangle_0 = v(1,1,1)^T$ \cite{s3} leads to the following form of the charged lepton mass matrix,

\begin{equation}
M_L = v \begin{bmatrix}
y_e & y_\mu & y_\tau\\
y_e & y_\mu \omega^2 & y_\tau \omega\\
y_e & y_\mu \omega & y_\tau \omega^2\\
\end{bmatrix}.
\end{equation}

The diagonalising matrices of $M_l$ are given by,

\begin{equation}
U_l = \frac{1}{\sqrt{3}} \begin{bmatrix}
1 & 1 & 1\\
1 & \omega^2 & \omega\\
1 & \omega & \omega^2\\                  
\end{bmatrix} \,\,\,\text{and} \,\,\,U_r=\begin{bmatrix}
1 & 0 & 0\\
0 & 1 & 0\\
0 & 0 & 1\\                  
\end{bmatrix}.
\end{equation}

The Dirac neutrino mass matrix is given by,

\begin{equation}
M_D =  \begin{bmatrix}
0 & d & 0\\
0 & 0 & d\\
d & 0 & 0\\
\end{bmatrix},
\end{equation}

where, $d=y_D \tilde{v}$

Assuming the VEVs of the fields $\tilde{\chi}$ and $\tilde{\kappa}$ along $v_{\chi}(1,1,1)^T$ and $v_{\kappa}(0,0,1)^T$ , the right handed neutrino mass matrix attains the following structure,

  \begin{equation}
M_R = \begin{bmatrix}
r & p & 0\\
p & r & 0\\
0 & 0 & r\\
\end{bmatrix},
\end{equation}

where, $r = y_{R_1} v_\chi$ and $p= y_{R_2} v_{\kappa}$.

Under the specific choice of the VEVs along the direction $\langle\Delta\rangle_0 = v_\Delta(0,-1,1)^T$, the contribution of Type-II mechanism to the neutrino mass matrix is obtained as shown below,

 \begin{equation}
M_{T_2} = \begin{bmatrix}
0 & t & -t\\
t & 0 & 0\\
-t & 0 & 0\\                  
\end{bmatrix}, \,\,\,\,\,\,\,\,\,\,\,\,\,\,\,\,\,\,\,\,\,\, \text{where},\,\,\,\, t=y_{T_2} v_\Delta
\end{equation}

The neutrino mass matrix is constructed after considering the contributions from Type-I and II see-saw mechanisms. After the charged lepton correction, the neutrino mass matrix is obtained as shown below,

\begin{equation}
M^f_\nu =
\begin{bmatrix}
A e^{i\alpha} & \textbf{B} e^{i\boldsymbol{\theta}}   &  \textbf{B} e^{-i\boldsymbol\theta} \\
B e^{i\theta}  &  C e^{i\beta} &  E e^{i\eta} \\
B e^{-i\theta} & E e^{i\eta} & D e^{i\gamma} \\
\end{bmatrix}.
\label{Type-P4 texture}
\end{equation}

From the matrix appearing in Eq.\ref{Type-P4 texture}, we can easily identify the relations of Type-p4 texture.

\subsection{Type-Q2}

The Type-Q2 texture shelters three constrained relations:

\begin{eqnarray}
2|m_{e\mu}|&=&|m_{\mu\mu}|\\
|m_{e\mu}|&=&|m_{e\tau}|\\
\text{arg}\,[m_{e\mu}]&=&-\text{arg}[m_{e\tau}]
\end{eqnarray}

 The Type-Q2 texture can be derived from the same Lagrangian appearing in Eq. \ref{Yukawa Lagrangian p4} by modifying the contribution to right-handed Majorana Mass term. This can be done in the following way,

\begin{equation}
\mathcal{L}_{Type-II}=\frac{1}{2}\,y_{R_1} (\bar{\nu}_{R}\,\nu^c_R)\chi.
\end{equation}

The right-handed Majorana Mass matrix is obtained as shown below,

  \begin{equation}
M_R = \begin{bmatrix}
r & 0 & 0\\
0 & r & 0\\
0 & 0 & r\\
\end{bmatrix},
\end{equation}

where, $r = y_{R_1} v_\chi$.

In flavour basis, the neutrino mass matrix is obtained as shown in the following,

\begin{equation}
M^f_\nu =
\begin{bmatrix}
A e^{i\alpha} & \textbf{B} e^{i\boldsymbol{\beta}}   &  \textbf{B} e^{-i\boldsymbol\beta} \\
B e^{i\beta}  &  \textbf{2B} e^{i\gamma} &  E e^{i\theta} \\
B e^{-i\beta} & E e^{i\eta} & D e^{i\psi} \\
\end{bmatrix}.
\label{Type-Q2 texture}
\end{equation}

\subsection{Type-R4}

The Type-R4 texture deals with two constrained relations: 

\begin{eqnarray}
|m_{ee}|&=&|m_{\mu\tau}|\\
|m_{e\mu}|&=&|m_{e\tau}|\\
|m_{\tau\tau}|&=&2|m_{e\mu}|\\
\text{arg}\,[m_{\mu\mu}]&=&-\text{arg}[m_{\tau\tau}]
\end{eqnarray}

 We try to derive this texture from $SU(2)_L \times A_4$ symmetry. In this regard, we expend the field content of the SM by adding right handed neutrinos, a singlet and a triplet scalar. The transformation properties of the field content are summarised in Table\ref{Field Chart of Type-R4}. We construct the $SU(2)_L \times A_4$ invariant Lagrangian in the following way,

\begin{eqnarray}
- \mathcal{L}_Y &=& y_e (\bar{D}_{l_L}\Phi)\,e_{R} + y_{\mu} (\bar{D}_{l_L}\Phi) \,\mu_{R}+y_{\tau} (\bar{D}_{l_L}\Phi)\, \tau_{R} + y_{D} \{(\bar{D}_{l_L}\tilde{\Phi})_{3_S}\, \nu_{R}+(\bar{D}_{l_L}\tilde{\Phi})_{3_A} \,\nu_{R}\}+\nonumber\\&&\frac{1}{2}\,y_{R_1} (\bar{\nu}_{R}\,\nu^c_R)_1 \,\eta \,+\, y_{T_2} (\bar{D}_{l_L}\, D_{l_L}^c)_3\,\Delta\,+\,h.c.,
\label{Yukawa Lagrangian R4}.
\end{eqnarray}

In symmetry the basis, the choice of the VEVs $(\Phi_i)$ as $\langle\Phi\rangle_{0}\,=\,v\,(1,1,1)^T$  lead us to the charged lepton mass matrix as shown below,

\begin{equation}
M_L = v \begin{bmatrix}
y_e & y_\mu & y_\tau\\
y_e & \omega^2 y_\mu & y_\tau \omega\\
y_e  & \omega y_\mu & y_\tau \omega^2\\
\end{bmatrix}.
\end{equation}

The left and right handed charged lepton diagonalizing matrices are of the obtained in the following manner,

 \begin{equation}
U_l = \frac{1}{\sqrt{3}} \begin{bmatrix}
1 & 1 & 1\\
1 & \omega^2 & \omega\\
1 & \omega & \omega^2\\                  
\end{bmatrix} \,\,\,\text{and} \,\,\,U_r=\begin{bmatrix}
1 & 0 & 0\\
0 & 1 & 0\\
0 & 0 & 1\\                  
\end{bmatrix}.
\end{equation}

The Dirac neutrino mass matrix is obtained as shown in the following,

\begin{equation}
M_D = 2\,y_{D} \,v \begin{bmatrix}
0 & 0 & 1\\
1 & 0 & 0\\
0 & 1 & 0\\
\end{bmatrix}.
\end{equation}

we assume the VEV of $\eta$ as $\langle\eta\rangle= v_\eta$ such that the right-handed neutrino mass matrix takes the following form,

   \begin{equation}
M_R =  \begin{bmatrix}
r & 0 & 0\\
0 & r & 0\\
0 & 0 & r\\
\end{bmatrix}.
\end{equation}

Assuming that $\Delta$ Higgs triplets take VEVs along the direction $\langle\Delta\rangle_0 = v_\Delta(0,-1,1)^T$, the Type-II contribution to the effective neutrino mass matrix takes the following form,

 \begin{equation}
M_{T_2} = d \begin{bmatrix}
0 & 1 & -1\\
1 & 0 & 0\\
-1 & 0 & 0\\                  
\end{bmatrix}, \,\,\,\,\,\,\,\,\,\,\,\,\,\,\,\,\,\,\,\,\,\, \text{where,}\,\,\,\,\,d=y_{T_2} v_\Delta
\end{equation}

In flavour basis, after taking the contribution from Type-I and Type-II see-saw mechanism, the neutrino mass matrix is obtained as shown in the following,

\begin{equation}
M^f_\nu =
\begin{bmatrix}
\textbf{A} e^{i\alpha} & \textbf{B} e^{i\beta}   &  \textbf{B} e^{-i\beta} \\
B e^{i\beta}  &  \textbf{2B} e^{i\boldsymbol\gamma} &  \textbf{A} e^{i\alpha} \\
B e^{-i\beta} &  A e^{i\alpha} & \textbf{2B} e^{-i\boldsymbol\gamma} \\
\end{bmatrix}.
\label{Type-R4 texture}
\end{equation}

From the matrix appearing in Eq.\ref{Type-R4 texture}, we identify the constrained relations under Type-R4 texture.

\section{Summary and Discussion}\label{Summary}

The Majorana CP phases are physical parameters. But the oscillation experiments cannot witness them. The neutrino mass matrix carries all the information regarding the masses, mixing, and CP phases. So it is quite relevant to draw constrained mass matrix textures that can predict the Majorana phases. In that light, we formulate forty-seven model independent mass matrix textures. The physics is independent of parametrization. But sometimes a suitable parametrization may help to understand the correlation between the different entities in a simple way. In this regard, we have emphasized on GP and EP. The Type-\textbf{A/P} textures can predict the Majorana phases $\alpha$ and $\beta$, while the Type-\textbf{Q/R} can predict $\alpha$, $\beta$ and $\delta$. Finally, the Type-\textbf{C/R} can predict $\theta_{23}$ in addition to $\alpha$, $\beta$ and $\delta$. This is to be emphasised that the present work deals with normal ordering of neutrino masses. However, a similar approach can be adopted for inverted and quasi-degenerate models.

In the present work, we follow a bottom up approach and try to set several promising textures of neutrino mass matrix by emphasizing on simple linear correlations among the different parameters. In principle, the neutrino mass matrix originates from the Yukawa Lagrangian, which can be constructed from Dimension-5 operator or seesaw models. The Higgs doublet/triplets may also contribute towards the effective neutrino mass matrix. Every neutrino mass matrix element carry the informations about the Yukawa couplings, the VEVs of Higgs scalar doublets or triplets and VEVs of different scalar fields. Now, if there exist some linear correlations among the mass matrix elements, then it is easy to trace out simple relations between the Yukawa couplings and that between the VEVs. This on the other hand may help the model builders to find out certain symmetry operations/Discrete groups under which the above relation could be relevant. In the present work, we have shown that the proposed textures (Type-A3, Type-B8, Type-C2, Type-P4, Type-Q2 and Type-R4) can be obtained starting from certain symmetry groups and attempt to realise the other textures is in progress. 

The textures elaborated in the present work may help the model builders to understand the first principle in a better way, and as a part of future work, we propose to study the stability of the proposed textures under the renormalization group running and to study the baryon asymmetry based upon the textures obtained in our work. Further, we can extend our work in $3+1$ neutrino mixing scenario with larger parameters space.

\section{Acknowledgement}

The research work of Pralay Chakraborty is supported by Innovation in Science Pursuit for Inspired Research (INSPIRE), Department of Science and Technology, Government of India, New Delhi vide grant No. IF190651. Manash Dey acknowledges financial support from the
Council of Scientific and Industrial Research (CSIR), Government of India through a NET Junior Research Fellowship vide grant No. 09/0059(15346)/2022-EMR-I.

\biboptions{sort&compress}

\appendix

\FloatBarrier
\section{Tables}

\newcolumntype{P}[1]{>{\centering\arraybackslash}p{#1}}
\begin{table}[h]
\centering
\begin{tabular}{P{2cm}P{4cm}} 
\hline\hline
S/N & ($m_1\,+\,m_2\,+\,m_3)/eV$ \\ 
\hline\hline
1 &  <\,0.26 \,~\cite{r1}\\
\hline
2 & <\,0.18 \,~\cite{r2}\\
\hline
3 & <\,0.152\,~\cite{r3}\\
\hline
4 & <\,0.14 \,~\cite{r4}\\
\hline
\end{tabular}
\caption{Upper bounds of the sum of three neutrino masses from Cosmological data.} 
\label{table:upper bound of sum of neutrino masses}
\end{table}

\begin{table}[h]
\centering
\begin{tabular}{P{2.5cm}P{2.5cm}P{2.5cm}} 
\hline\hline
Parameters & btf\,$\pm$\,$1\,\sigma$ & $3\,\sigma$\,range\\ 
\hline\hline
$\theta_{12}/^{\circ}$ & $33.44^{+0.78}_{-0.75}$ & 31.27\,-\,35.86\\
\hline
$\theta_{23}/^{\circ}$ & $49^{+1.1}_{-1.4}$ & 39.6\,-\,51.8\\
\hline
$\theta_{13}/^{\circ}$ & $8.57^{+0.13}_{-0.12}$ & 8.20\,-\,8.97\\
\hline
$\delta_{CP}/^{\circ}$ & $195^{+51}_{-25}$ & 107\,-\,403\\
\hline
$\frac{\Delta\,m_{21}^2}{10^{-5}\,eV^2}$ & $7.42^{+0.21}_{-0.20}$ & 6.82\,-\,8.04\\
\hline
$\frac{|\Delta\,m_{31}^2|}{10^{-3}\,eV^2}$ & $2.515^{+0.028}_{-0.027}$ & 2.431\,-\,2.598\\
\hline
\end{tabular}
\caption{$3\,\nu$ oscillation parameters obtained from different global analysis of neutrino data in normal ordering\,~\cite{n}.} 
\label{table:Values Of Parameters in NO}
\end{table}

\begin{table}[h]
\centering
\begin{tabular}{P{1cm}P{2cm}P{4.3cm}P{1cm}P{2cm}P{4.2cm}} 
\hline\hline
S/N & Texture & Conditions & S/N & Texture & Conditions\\ 
\hline\hline
1 & Type-A1 & $\,\text{Re}[m_{ee}]=-\text{Im}\,[m_{e\mu}]\linebreak
           \text{Re}\,[m_{e\mu}]=-\text{Im}\,[m_{e\tau}]$ & 6 & Type-A6 & $\text{Re}\,[m_{e\mu}]=-\text{Im}\,[m_{\mu\mu}]\linebreak
           \text{Im}\,[m_{\mu\mu}]=\text{Im}\,[m_{e\mu}]$\\ 
\hline
2 & Type-A2 & $\text{Im}\,[m_{ee}]=-\text{Re}\,[m_{e\tau}]\linebreak
           \text{Im}\,[m_{e\mu}]=\text{Im}\,[m_{e\tau}]$ & 7 & Type-A7 & $\text{Re}\,[m_{e\mu}] = -\text{Im}\,[m_{\mu\tau}]\linebreak
           \text{Re}\,[m_{e\tau}]=-\text{Im}\,[m_{\tau\tau}]$\\ 
\hline
3 & Type-A3 & $\text{Re}\,[m_{\tau\tau}]=-2\,\text{Re}\,[m_{e\tau}]\linebreak
           \text{Im}\,[m_{e\tau}]=\text{Im}\,[m_{\mu\tau}]$ & 8 & Type-A8 & $\text{Re}\,[m_{ee}]=-\text{Im}\,[m_{\mu\tau}]\linebreak
           \text{Re}\,[m_{e\mu}]=-\text{Im}\,[m_{\tau\tau}]$ \\
\hline
4 & Type-A4 & $\text{Re}\,[m_{\mu\mu}]=\text{Re}\,[m_{\mu\tau}]\linebreak
           \text{Re}\,[m_{e\tau}]=-\text{Im}\,[m_{e\tau}]$ & 6 &  Type-A9 & $\text{Re}\,[m_{e\mu}]=-\text{Im}\,[m_{\tau\tau}]\linebreak
           \text{Re}\,[m_{e\tau}]=-\text{Im}\,[m_{\mu\tau}]$ \\ 
\hline
5 & Type-A5 & $\text{Re}\,[m_{ee}]=-\text{Im}\,[m_{\tau\tau}]\linebreak
           \text{Im}\,[m_{\mu\tau}]=\text{Im}\,[m_{e\mu}]$ & 10 &  Type-A10 & $\text{Re}\,[m_{\mu\mu}]=\text{Re}\,[m_{\mu\tau}]\linebreak
           \text{Re}\,[m_{e\mu}]=-\text{Im}\,[m_{e\mu}]$\\ 
\hline

11 & Type-A11 & $\text{Re}\,[m_{ee}]=-\text{Im}\,[m_{\mu\mu}]\linebreak
           \text{Re}\,[m_{e\mu}]=-\text{Im}\,[m_{\mu\tau}]$ &  &  &  \\
\hline
\end{tabular}
\caption{Conditions of Type-A Textures.} 
\label{table:Type-A textures}
\end{table}

\begin{table}[!]
\centering
\begin{tabular}{P{1cm}P{2cm}P{4.5cm}P{1cm}P{2cm}P{4.5cm}} 
\hline\hline
S/N & Texture & Conditions & S/N & Texture  & Conditions \\ 
\hline\hline
\vspace{.2cm}1 & \vspace{.05cm}Type-B1 & 
$\text{Im}\,[m_{ee}]=-3 \,\text{Re}\,[m_{ee}]\linebreak
           \text{Re}\,[m_{e\mu}]=-\text{Im}\,[m_{e\mu}]\linebreak
           2 \,\text{Re}\,[m_{e\tau}]=-\text{Im}\,[m_{e\tau}]$
& \vspace{.2cm} 7 & \vspace{.1cm}Type-B7 & $\text{Re}\,[m_{ee}]=-3 \,\text{Im}\,[m_{\tau\tau}]\linebreak
           2\, \text{Re}\,[m_{ee}]= -\text{Im}\,[m_{\mu\tau}]\linebreak
           \text{Im}\,[m_{\mu\mu}]=2 \,\text{Im}\,[m_{\tau\tau}]$\\ 
\hline
\vspace{.2cm} 2 & \vspace{.2cm}Type-B2 & $\text{Im}\,[m_{ee}]= -\text{Re}\,[m_{e\mu}]\linebreak
           \text{Re}\,[m_{e\tau}]= -2\, \text{Im}\,[m_{e\mu}]\linebreak
           \text{Re}\,[m_{ee}]= -3 \,\text{Im}\,[m_{e\tau}]$ & \vspace{.2cm} 8 & \vspace{.2cm}Type-B8 & $\text{Re}\,[m_{e\mu}]= \text{Im}\,[m_{e\tau}]\linebreak
           \text{Re}\,[m_{ee}]= - \,\text{Im}[m_{\mu\tau}]\linebreak
           \text{Im}\,[m_{\mu\mu}]= - \,\text{Im}[m_{\tau\tau}]$\\ 
\hline
\vspace{.2cm} 3 & \vspace{.2cm}Type-B3 & $\text{Re}\,[m_{\mu\mu}]= \text{Re}\,[m_{\mu\tau}]\linebreak
           \text{Re}\,[m_{\mu\tau}]= \text{Re}\,[m_{\tau\tau}]\linebreak
           2 \,\text{Re}\,[m_{e\mu}]= - \text{Im}\,[m_{\tau\tau}]$ & \vspace{.2cm} 9 & \vspace{.2cm}Type-B9 & $\text{Im}\,[m_{e\mu}]=\text{Im}\,[m_{\mu\tau}]\linebreak
           \text{Im}\,[m_{\mu\tau}]= -\,\text{Im}\,[m_{\tau\tau}]\linebreak
           \text{Re}\,[m_{ee}]=\,2 \text{Im}[m_{\mu\mu}]$\\ 
\hline
\vspace{.2cm} 4 & \vspace{.2cm}Type-B4 & $\text{Re}\,[m_{e\mu}]=-\text{Im}\,[m_{\mu\mu}]\linebreak
           \text{Im}\,[m_{e\mu}]= 2\,\text{Im}\,[m_{\tau\tau}]\linebreak
           \text{Re}\,[m_{e\tau}]=-\text{Im}\,[m_{\mu\tau}]$ & \vspace{.2cm} 10 & \vspace{.2cm}Type-B10 & $\text{Re}\,[m_{e\tau}]= -\, \text{Re}\,[m_{e\mu}]\linebreak
           \text{Im}\,[m_{e\mu}]= - \,\text{Im}\,[m_{\tau\tau}]\linebreak
           \text{Im}\,[m_{ee}]= -\,\text{Im}\,[m_{\mu\tau}]$\\ 
\hline
\vspace{.2cm}5 & \vspace{.2cm}Type-B5 & $\text{Re}\,[m_{\mu\mu}]=-3\,\text{Im}\,[m_{\mu\tau}]\linebreak
           2\, \text{Re}\,[m_{e\mu}]= \text{Re}\,[m_{e\tau}]\linebreak
           3\, \text{Re}\,[m_{e\tau}]=-\text{Im}\,[m_{\mu\mu}]$ & \vspace{.2cm} 11 & \vspace{.2cm}Type-B11 & $\text{Im}\,[m_{e\tau}]= - \,\text{Im}\,[m_{ee}]\linebreak
           3 \,\text{Re}\,[m_{e\mu}]= -\,\text{Re}\,[m_{\mu\mu}]\linebreak
           \text{Im}\,[m_{\mu\mu}]=- \,\text{Im}\,[m_{\tau\tau}]$\\ 
\hline
\vspace{.2cm} 6 & \vspace{.2cm}Type-B6 & $ 2\,\text{Re}\,[m_{e\tau}]=\text{Re}\,[m_{\mu\mu}]\linebreak
           3\,\text{Re}\,[m_{e\tau}]= \text{Re}\,[m_{\mu\tau}]\linebreak
           \text{Im}\,[m_{\tau\tau}]=\text{Im}\,[m_{\mu\mu}]$ & \vspace{.2cm}12 & \vspace{.2cm} Type-B12 & $\text{Im}\,[m_{ee}]=- \,\text{Im}\,[m_{\tau\tau}]\linebreak
           \text{Re}\,[m_{e\mu}]= - \,\text{Re}[m_{e\tau}]\linebreak
           \text{Re}\,[m_{ee}]= \,2 \,\text{Im}[m_{\mu\tau}]$\\ 
\hline
\end{tabular}
\caption{Conditions of Type-B Textures.}
\label{table:Type-B textures}
\end{table}

\begin{table}[!]
\centering
\begin{tabular}{P{1cm}P{2cm}P{4.5cm}P{1cm}P{2cm}P{4.5cm}} 
\hline\hline
S/N & Texture  & Conditions & S/N & Texture  & Conditions \\ 
\hline\hline
\vspace{.5cm} 1 & \vspace{.2cm}Type-C1 & $3\, \text{Im}\,[m_{\mu\mu}]= \text{Im}\,[m_{e\tau}]\linebreak
           \text{Re}\,[m_{e\mu}]=-\text{Re}\,[m_{e\tau}]\linebreak
           \text{Im}\,[m_{e\tau}]=-\text{Re}\,[m_{ee}]\linebreak
           \text{Im}\,[m_{\tau\tau}]= 2\,\text{Im}\,[m_{\mu\tau}]$ & \vspace{.5cm} 4& \vspace{.2cm}Type-C4 & $\text{Re}\,[m_{\mu\mu}]=\text{Re}\,[m_{\tau\tau}]\linebreak
           3 \,\text{Re}\,[m_{e\tau}]=-\text{Im}\,[m_{e\mu}]\linebreak
           \text{Re}\,[m_{e\tau}]=-\text{Im}\,[m_{\tau\tau}]\linebreak
           \text{Im}\,[m_{e\mu}]= 3\,\text{Im}\,[m_{\mu\mu}]$ \\ 
\hline
\vspace{.5cm}2 & \vspace{.2cm}Type-C2 & $\text{Re}\,[m_{ee}]= \text{Re}\,[m_{\mu\tau}]\linebreak
           \text{Re}\,[m_{e\mu}]= - \,\text{Re}\,[m_{e\tau}]\linebreak
           \text{Im}\,[m_{\mu\mu}]= \text{Im}\,[m_{\tau\tau}]\linebreak
           \text{Im}\,[m_{\mu\mu}]= -\,\text{Im}\,[m_{\mu\tau}]$ & \vspace{.5cm}5 & \vspace{.2cm}Type-C5 & $\text{Re}\,[m_{ee}]=- 3 \,\text{Im}\,[m_{ee}]\linebreak
           \text{Re}\,[m_{e\mu}]= -2 \,\text{Im}\,[m_{e\tau}]\linebreak
           \text{Re}\,[m_{e\tau}]=- 3\,\text{Im}\,[m_{ee}]\linebreak
           \text{Re}\,[m_{\mu\tau}]= \text{Re}\,[m_{\tau\tau}]$ \\ 
\hline
\vspace{.5cm}3 & \vspace{.2cm}Type-C3 & $\text{Re}\,[m_{ee}]=-\text{Im}\,[m_{\mu\mu}]\linebreak
           \text{Re}\,[m_{e\mu}]=-\text{Im}\,[m_{\mu\tau}]\linebreak
           \text{Re}\,[m_{\mu\mu}]=\text{Re}\,[m_{\mu\tau}]\linebreak
           \text{Im}\,[m_{e\tau}]= 2 \,\text{Im}\,[m_{e\mu}]$
\\ 
\hline
\end{tabular}
\caption{Conditions of Type-C Textures.}
\label{table:Type-C textures}
\end{table}

\begin{table}[!]
\centering
\begin{tabular}{P{1cm}P{2cm}P{4.5cm}P{1cm}P{2cm}P{4.5cm}} 
\hline\hline
S/N & Texture & Conditions & S/N & Texture & Conditions \\ 
\hline\hline
1 & Type-P1 & $2 \,|m_{ee}| = |m_{\mu\mu}|\linebreak
           |m_{e\mu}|=|m_{e\tau}|$ & 7 & Type-P7 & $|m_{\mu\tau}|=2\,\text{arg}\,[m_{\mu\mu}]\linebreak
           |m_{ee}|=\text{arg}\,[m_{\mu\mu}]$\\ 
\hline
2 & Type-P2 & $|m_{\mu\mu}|=|m_{\tau\tau}|\linebreak
           |m_{e\mu}|=2|m_{ee}|$ & 8 & Type-P8 & $|m_{\tau\tau}|=\,3\,\text{arg}\,[m_{\mu\mu}]\linebreak
           |m_{ee}|=2\, \text{arg}\,[m_{e\mu}]$\\ 
\hline
3 & Type-P3 &  $2\,|m_{e\tau}|=|m_{\tau\tau}|\linebreak
           2\,|m_{e\mu}|=|m_{\mu\mu}|$ & 9 & Type-P9 & $|m_{e\tau}|=\text{arg}\,[m_{\mu\mu}]\linebreak
           |m_{ee}|= 3\, \text{arg}\,[m_{\tau\tau}]$\\
\hline
4 & Type-P4 &  $|m_{e\mu}|=|m_{e\tau}|\linebreak
           \text{arg}[m_{e\mu}]=-\text{arg}|m_{e\tau}|$ & 10 & Type-P10 &  $|m_{e\tau}|=\text{arg}\,|m_{e\tau}|\linebreak
           |m_{ee}|=2\,\text{arg}\,[m_{e\tau}]$\\ 
\hline
5 & Type-P5 & $|m_{\mu\tau}|=2\,|m_{e\tau}|\linebreak
           2\,|m_{e\mu}|=|m_{\tau\tau}|$ & 11 & Type-P11 &  $2 \,\text{arg}[m_{\mu\tau}]=|m_{\mu\tau}|\linebreak
           \text{arg}\,[m_{\mu\tau}]=2\,|m_{e\tau}|$\\ 
\hline
6 & Type-P6 & $\text{arg}\,[m_{\mu\mu}]=\text{arg}\,[m_{\tau\tau}]\linebreak
           \text{arg}\,[m_{e\tau}]=\text{arg}\,[m_{\mu\tau}]$ & 12 & Type-P12 & $2\, |m_{\mu\tau}|=|m_{\tau\tau}|\linebreak
           2\,|m_{e\tau}|=|m_{\mu\mu}|$\\  
\hline
\end{tabular}
\caption{Conditions of Type-P Textures.}
\label{table:Type-P textures} 
\end{table}

\begin{table}[!]
\centering
\begin{tabular}{P{1cm}P{2cm}P{4.5cm}P{1cm}P{2cm}P{4.5cm}} 
\hline\hline
S/N & Texture & Conditions & S/N & Texture & Conditions \\ 
\hline\hline
\vspace{.2cm} 1 & \vspace{.2cm}Type-Q1 &  $3\,|m_{ee}|= \text{arg}\,[m_{e\mu}]\linebreak
           \text{arg}\,[m_{\tau\tau}]= 2\, |m_{e\mu}|\linebreak
           |m_{e\tau}|=2\, \text{arg}\,[m_{\mu\mu}]$ & \vspace{.2cm} 5 & \vspace{.2cm} Type-Q5 & $|m_{\mu\mu}|=2\,|m_{ee}|\linebreak
            |m_{e\mu}|=|m_{ee}|\linebreak
          \text{arg}\,[m_{\mu\tau}]=|m_{e\tau}|$ \\ 
\hline
\vspace{.2cm}2 & \vspace{.2cm}Type-Q2 &  $2|m_{e\mu}|= |m_{\mu\mu}|\linebreak
            |m_{e\mu}|=|m_{e\tau}|\linebreak
           \text{arg}[m_{e\mu}]=-\text{arg}[m_{e\tau}]$ & \vspace{.2cm}6 & \vspace{.2cm}Type-Q6 & $\text{arg}\,[m_{\mu\tau}]=\text{arg}\,[m_{e\tau}]\linebreak
          \text{arg}\,[m_{e\tau}]=2\,\text{arg}\,[m_{\tau\tau}]\linebreak
           |m_{\mu\mu}|=2\,|m_{ee}|$\\ 
\hline
\vspace{.2cm}3 & \vspace{.2cm}Type-Q3 &  $\text{arg}\,[m_{\mu\mu}]= 3\,\text{arg}\,[m_{\tau\tau}]\linebreak
            |m_{e\mu}|=\text{arg}\,[m_{e\tau}]\linebreak
           |m_{e\tau}|=\text{arg}\,[m_{\tau\tau}]$ & \vspace{.2cm}7 & \vspace{.2cm}Type-Q7 & $2\,|m_{e\tau}|=\text{arg}\,[m_{e\tau}]\linebreak
            3 \,\text{arg}\,[m_{\tau\tau}]=|m_{ee}|\linebreak
           4\, \text{arg}[m_{\mu\tau}]=|m_{\tau\tau}|$\\  
\hline
\vspace{.2cm}4 & \vspace{.2cm}Type-Q4 &  $\text{arg}\,[m_{\mu\tau}]=|m_{\mu\tau}|\linebreak
            \text{arg}\,[m_{\mu\tau}]=|m_{e\tau}|\linebreak
           \text{arg}\,[m_{e\mu}]=\text{arg}\,[m_{e\tau}]$ & \vspace{.2cm} & \vspace{.2cm}  & \vspace{.2cm} \\ 
\hline
\end{tabular}
\caption{Conditions of Type-Q Textures.}
\label{table:Type-Q textures} 
\end{table}

\begin{table}[!]
\centering
\begin{tabular}{P{1cm}P{2cm}P{4.5cm}P{1cm}P{2cm}P{4.5cm}} 
\hline\hline
S/N & Texture & Conditions & S/N & Texture & Conditions \\ 
\hline\hline
\vspace{.5cm}1 & \vspace{.2cm}Type-R1 & $|m_{ee}|=\text{arg}\,[m_{\tau\tau}]\linebreak
            2\, \text{arg}\,[m_{\tau\tau}]=|m_{e\mu}|\linebreak
            2\,\text{arg}\,[m_{ee}]=|m_{e\tau}|\linebreak
          |m_{\mu\tau}|=3\,\text{arg}\,[m_{ee}]$ & \vspace{.5cm}4 & \vspace{.5cm}Type-R4 & $|m_{ee}|=|m_{\mu\tau}|\linebreak
            |m_{e\mu}|=|m_{e\tau}|\linebreak
            |m_{\tau\tau}]=2\,|m_{e\mu}|\linebreak
           \text{arg}\,|m_{\mu\mu}|=-\text{arg}\,[m_{\tau\tau}]$\\ 
\hline
\vspace{.5cm}2 & \vspace{.2cm}Type-R2 & $\,|m_{e\mu}|=2\,\text{arg}\,[m_{\tau\tau}]\linebreak
            3\,|m_{ee}|=\text{arg}\,[m_{ee}]\linebreak
            \text{arg}\,[m_{ee}]=\text{arg}[m_{e\tau}]\linebreak
          |m_{\mu\tau}|=\,\text{arg}\,[m_{ee}]$ & \vspace{.5cm}5 & \vspace{.2cm}Type-R5 & $|m_{\mu\tau}|=2\,\text{arg}\,[m_{e\tau}]\linebreak
            \text{arg}\,[m_{\mu\mu}]=3\,|m_{e\tau}|\linebreak
            |m_{ee}|=2\,\text{arg}[m_{e\tau}]\linebreak
           |m_{\mu\mu}|=2\,\text{arg}\,[m_{\tau\tau}]$\\ 
\hline
\vspace{.5cm}3 & \vspace{.2cm}Type-R3 & $\text{arg}\,[m_{\mu\mu}]=3\,\text{arg}\,[m_{\tau\tau}]\linebreak
            |m_{e\mu}|=\text{arg}\,[m_{e\tau}]\linebreak
            |m_{e\tau}|=\text{arg}\,[m_{\tau\tau}]\linebreak
           2\,\text{arg}\,[m_{\mu\tau}]=|m_{e\tau}|$ & \vspace{.5cm}6 & \vspace{.5cm}Type-R6 & $3\,|m_{\mu\mu}|=\text{arg}\,[m_{ee}]\linebreak
            |m_{ee}|=2\,|m_{e\mu}|\linebreak
            \text{arg}\,[m_{\mu\mu}]=\text{arg}\,[m_{\tau\tau}]\linebreak
           2\,|m_{e\mu}|=\text{arg}\,[m_{e\mu}]$ \\ 
\hline
\end{tabular}
\caption{Conditions of Type-R Textures.}
\label{table:Type-R textures} 
\end{table}

\begin{table}[!]
\centering
\begin{tabular}{c c c c c c } 
\hline
\hline
& & \multicolumn{2}{c}{Initial value} & \multicolumn{2}{c}{Estimated value} \\ 
S/N & Texture & $\alpha_0/^{\circ}$  & $\beta_0/^{\circ}$ & $\alpha^{\circ}$  & $\beta/^{\circ}$ \\ 
\hline
\hline
1 & Type-A1 & 50.00  & 60.00 & 47.86 &  52.44\\
\hline
2 & Type-A2 & 57.29 & 68.75 & 90.21 & 90.71 \\ 
\hline
3 & Type-A3 & 90.00 & 60.00 & 97.24 & 44.88 \\
\hline
4 & Type-A4 & 68.75 & 39.53 & 51.55  & 47.79 \\
\hline
5 & Type-A5 & 68.75 & 70.00 & 39.89 & 91.36 \\ 
\hline
6 & Type-A6 & 45.00 & 45.00 & 23.62 & 48.97 \\ 
\hline
7 & Type-A7 & 50.00 & 80.00 & 88.97 & 91.30 \\ 
\hline
8 & Type-A8 & 28.65 & 68.75 & 15.71 & 63.73 \\ 
\hline
9 & Type-A9 & 10.00 & 65.00 & 37.82 & 75.56 \\ 
\hline
10 & Type-A10 & 15.00 & 30.00 & 42.96 & 51.59 \\ 
\hline
11 & Type-A11 & 90.00 & 45.00 & 72.81 & 24.69 \\
\hline
\end{tabular}
\caption{Estimated values of $\alpha$ and $\beta$ along with the initial values of Type-A Textures. To predict the Majorana phases for the textures of Type-A, the best fit values of the  mixing angles and mass squared differences are taken as input ~\cite{n}. The input value of $\delta$ is fixed at $270^\circ$. The $m_3$ is chosen to be $0.06$ eV. }
\label{table:Values of Type-A textures}
\end{table}
\begin{table}[!]
\centering
\begin{tabular}{c c c c c c c c} 
\hline
\hline
& & \multicolumn{3}{c}{Initial value} & \multicolumn{3}{c}{Estimated value}\\
S/N & Texture &  $\alpha_0/^{\circ}$  &  $\beta_0/^{\circ}$ & $\delta_0/^{\circ}$ & $\alpha/^{\circ}$  & $\beta/^{\circ}$ & $\delta/^{\circ}$\\ 
\hline\hline
1 & Type-B1 & 50.00 & 50.00 & 220 & 51.93 & 55.51 & 259.99 \\
\hline
2 & Type-B2 & 108.86 & 148.97 & 343.77 & 114.31 & 161.40 & 348.21\\ 
\hline
3 & Type-B3 & 126.05 & 40.11 & 229.18 & 137.30 &45.62 & 240.92\\  
\hline 
4 & Type-B4 & 90.00 & 45.00 & 229.18 & 85.84 & 90.41 & 272.22\\ 
\hline
5 & Type-B5 & 51.57 & 45.84 & 229.18 & 29.41 & 26.27 & 270.73\\ 
\hline
6 & Type-B6 & 70.00 & 80.00 & 220.00 & 32.15 & 72.50 & 134.07\\ 
\hline
7 & Type-B7 & 70.00 & 45.00 & 180.00 & 122.61 & 11.27 & 136.49\\ 
\hline
8 & Type-B8 & 105.00 & 90.00 & 120.00 & 147.32 & 79.00 & 297.58\\ 
\hline
9 & Type-B9 & 70.00 & 90.00 & 210.00 & 16.54 & 84.93 & 363.25\\ 
\hline
10 & Type-B10 & 90.00 & 80.00 & 190.00 & 90.28 & 93.74 & 268.92\\ 
\hline
11 & Type-B11 & 60.00 & 70.00 & 120.00 & 89.33 & 91.12 & 125.25\\
\hline
12 & Type-B12 & 45.00 & 90.00 & 229.00 & 10.52 & 108.02 & 257.11\\ 
\hline
\end{tabular}
\caption{Estimated values of $\alpha$, $\beta$ and $\delta$ along with the initial values of Type-B Textures. To predict the phases $\alpha$, $\beta$ and $\delta$ for the textures of Type-B, the best fit values of the  mixing angles and mass squared differences are taken as input ~\cite{n}. The $m_3$ is fixed at $0.06$ eV.}
\label{table:Values of Type-B textures}
\end{table}

\begin{table}[!]
\centering
\begin{tabular}{c c c c c c c c c c}
\hline
\hline
&  &  \multicolumn{4}{c}{Initial value} & \multicolumn{4}{c}{Estimated value} \\ 
S/N & Texture & $\alpha_0/^{\circ}$ & $\beta_0/^{\circ}$& $\delta_0/^{\circ}$ & $\theta_{23_0}/^{\circ}$ & $\alpha/^{\circ}$ & $\beta/^{\circ}$& $\delta/^{\circ}$ & $\theta_{23}/^{\circ}$\\
\hline\hline   
1 & Type-C1 & 40.00 & 98.00 & 105.00 & 40.00 & 41.41 & 100.30 &  99.81 & 41.16\\ 
\hline
2 & Type-C2 & 10.00 & 120.00 & 290.00 & 49.00 & 17.27 & 144.55 & 270.00 & 48.75\\ 
\hline
3 & Type-C3 & 154.70 & 114.59 & 229.18 & 41.83 & 166.06 & 113.56 & 301.08 & 48.77\\ 
\hline
4 & Type-C4 & 126.05 & 45.84 & 229.18 & 41.83 & 116.086 & 37.86 & 281.79 & 49.19\\ 
\hline
5 & Type-C5 & 120.00 & 20.00 & 120.00 & 40.00 & 93.49 & 28.19 & 276.36 & 51.00\\ 
\hline
\end{tabular}
\caption{Estimated values of $\alpha$, $\beta$, $\delta$ and $\theta_{23}$ along with the initial values of Type-C Textures. To predict the said parameters, we seed the best fit values of the  mixing angles ($\theta_{12}$ and $\theta_{13}$) and mass squared differences as input~\cite{n}. The $m_3$ is fixed at $0.06$ eV.}
\label{table:Values of Type-C textures}
\end{table}

\begin{table}[!]
\centering
\begin{tabular}{c c c c c c } 
\hline
\hline
& & \multicolumn{2}{c}{Initial value} & \multicolumn{2}{c}{Estimated value} \\ 
S/N & Texture & $\alpha_0/^{\circ}$  & $\beta_0/^{\circ}$ & $\alpha^{\circ}$  & $\beta/^{\circ}$ \\ 
\hline
\hline
1 & Type-P1 & 50.00 & 100.00 & 6.77 & 80.77\\ 
\hline
2 & Type-P2 & 160.00 & 63.00 & 163.91 & 60.36\\ 
\hline
3 & Type-P3 & 95.74 & 74.08 & 99.74 & 81.79\\
\hline
4 & Type-P4 & 12.00 & 167.00 &  12.38 & 79.51\\ 
\hline
5 & Type-P5 & 85.20 & 46.29 &  77.40 & 34.62\\ 
\hline
6 & Type-P6 & 100.67 & 87.72 &  94.98 & 88.38\\ 
\hline
7 & Type-P7 & 60.00 & 57.29 & 201.87 & 129.58 \\
\hline
8 & Type-P8 & 57.29 & 85.94 & 40.09 & 87.78\\
\hline
9 & Type-P9 & 81.53 & 87.66 &  88.49 & 88.16\\ 
\hline
10 & Type-P10 & 158.14 & 111.78 & 143.28 & 113.48\\
\hline
11 & Type-P11 & 130.29 & 111.78 & 131.61 & 111.73 \\ 
\hline
12 & Type-P12 & 60.00 & 10.00 & 46.72 & 0.55\\ 
\hline
\end{tabular}
\caption{Estimated and initial values of $\alpha$ and $\beta$ along with the input value of $\delta$ for Type-P Textures. To predict the said parameters, the best fit values of the  mixing angles and mass squared differences are taken as input ~\cite{n}. The input of $\delta$ is set at $270^\circ$ whereas the $m_3$ is fixed at $0.06$ eV.}
\label{table:Values of Type-P textures}
\end{table}

\begin{table}[!]
\centering
\begin{tabular}{c c c c c c c c} 
\hline
\hline
& & \multicolumn{3}{c}{Initial value} & \multicolumn{3}{c}{Estimated value}\\
S/N & Texture &  $\alpha_0/^{\circ}$  &  $\beta_0/^{\circ}$ & $\delta_0/^{\circ}$ & $\alpha/^{\circ}$  & $\beta/^{\circ}$ & $\delta/^{\circ}$\\ 
\hline\hline  
1 & Type-Q1 & 110.00 & 60.00 & 200.00 & 92.57 & 26.88 & 165.01  \\ 
\hline
2 & Type-Q2 & 20.00 & 65.00 & 120.00 & 23.35 & 66.47 & 125.33 \\ 
\hline
3 & Type-Q3 & 84.80 & 91.67 & 120.00 & 85.42 & 91.87 & 95.78 \\ 
\hline
4 & Type-Q4 & 17.19 & 166.16 & 200.53 & 1.93 & 172.97 & 264.93 \\ 
\hline
5 & Type-Q5 & 10.00 & 148.00 & 266.42 & 77.68 & 194.12 & 287.33 \\ 
\hline
6 & Type-Q6 & 165.81 & 91.96 & 143.24 & 166.27 & 91.96 & 276.43 \\ 
\hline
7 & Type-Q7 & 45.84 & 97.40 & 171.89 & 33.14 & 94.24 & 126.84 \\ 
\hline
\end{tabular}
\caption{Estimated values of $\alpha$, $\beta$ and $\delta$ along with the initial values of Type-Q Textures. To predict the said phases, we take the best fit values of the  mixing angles and mass squared differences as input ~\cite{n}. The $m_3$ is fixed at $0.06$ eV.}
\label{table:Values of Type-Q textures}
\end{table}

\begin{table}[!]
\centering
\begin{tabular}{c c c c c c c c c c}
\hline
\hline
&  &  \multicolumn{4}{c}{Initial value} & \multicolumn{4}{c}{Estimated value} \\ 
S/N & Texture & $\alpha_0/^{\circ}$ & $\beta_0/^{\circ}$& $\delta_0/^{\circ}$ & $\theta_{23_0}/^{\circ}$ & $\alpha/^{\circ}$ & $\beta/^{\circ}$& $\delta/^{\circ}$ & $\theta_{23}/^{\circ}$\\
\hline\hline
1 & Type-R1 &70.00 & 10.00 & 300.00 & 43.00 & 104.02 & 23.11 & 313.05 & 51.55 \\ 
\hline
2 & Type-R2 & 11.26 & 90.62 & 231.98 & 43.10 & 10.47 & 90.77 & 247.11 & 41.52 \\ 
\hline
3 & Type-R3 & 114.59 & 91.67 & 194.81 & 42.97 &  95.74 & 96.13 & 160.74 & 41.70\\ 
\hline
4 & Type-R4 & 48.94 & 167.93 & 332.10 & 40.10 & 50.82 & 167.85 & 322.76 & 45.53 \\ 
\hline
5 & Type-R5 & 85.94 & 68.75 & 194.81 & 42.97 & 60.49 & 18.97 & 10.74 & 44.52 \\
\hline
6 & Type-R6 & 80.21 & 80.21 & 270.00 & 42.00 & 26.25 & 75.03 & 353.75 & 43.57 \\ 
\hline
\end{tabular}
\caption{Estimated values of $\alpha$, $\beta$, $\delta$ and $\theta_{23}$ along with the initial values of Type-R Textures. To predict the said parameters, we put the best fit values of the  mixing angles ($\theta_{12}$ and $\theta_{13}$) and mass squared differences as input ~\cite{n}. The $m_3$ is chosen to be $0.06$ eV.}
\label{table:Values of Type-R textures} 
\end{table}

\begin{table}[!]
\centering
\begin{tabular}{P{1.8cm}P{1.5cm}P{1.8cm}P{1.5cm}P{1.8cm}P{1.5cm}P{1.8cm}P{1.5cm}} 
\hline\hline
Texture & $m_{\beta\beta}$ $(meV)$ & Texture & $m_{\beta\beta}$ $(meV)$ & Texture & $m_{\beta\beta}$ $(meV)$ & Texture & $m_{\beta\beta}$ $(meV)$ \\ 
\hline\hline
Type-A1 & 32.67 & Type-A8 & 22.90 & Type-B4 & 33.78 & Type-B11 & 32.99\\
\hline
Type-A2 & 33.87 & Type-A9  & 27.00 & Type-B5 & 31.81 & Type-B12 & 11.65\\
\hline
Type-A3 & 23.44 & Type-A10  & 32.27 & Type-B6 & 27.37 & Type-C1 & 20.71\\
\hline
Type-A4 & 32.75 & Type-A11  & 24.28 & Type-B7 & 15.15 & Type-C2 & 20.69\\
\hline
Type-A5 & 22.84 & Type-B1  & 31.30 & Type-B8 & 15.52 & Type-C3 & 20.77\\
\hline
Type-A6 & 29.28 & Type-B2  & 23.08 & Type-B9 & 17.194 & Type-C4 & 14.50\\
\hline
Type-A7 & 33.85 & Type-B3 & 13.40 & Type-B10 & 33.81 & Type-C5 & 18.93\\
\hline
\end{tabular}
\caption{Estimated values of $m_{\beta\beta}$ associated with textures under GP.}
\label{table:Values of DBB under GP}
\end{table}

\begin{table}[!]
\centering
\begin{tabular}{P{1.8cm}P{1.5cm}P{1.8cm}P{1.5cm}P{1.8cm}P{1.5cm}P{1.8cm}P{1.5cm}} 
\hline\hline
Texture & $m_{\beta\beta}$ $(meV)$ & Texture & $m_{\beta\beta}$ $(meV)$ & Texture & $m_{\beta\beta}$ $(meV)$ & Texture & $m_{\beta\beta}$ $(meV)$ \\ 
\hline\hline
Type-P1 & 13.78 & Type-P8 & 24.10 & Type-Q3 & 33.68 & Type-R3 & 31.35\\
\hline
Type-P2 & 12.90 & Type-P9  & 33.86 & Type-Q4 & 31.60 & Type-R4 & 19.03\\
\hline
Type-P3 & 32.50 & Type-P10  & 28.94 & Type-Q5 & 19.36 & Type-R5 & 24.18\\
\hline
Type-P4 & 16.16 & Type-P11  & 31.34 & Type-Q6 & 13.70 & Type-R6 & 25.60 \\
\hline
Type-P5 & 26.20 & Type-P12  & 23.77 & Type-Q7 & 20.37 &  &\\
\hline
Type-P6 & 33.67 & Type-Q1  & 16.66 & Type-R1 & 13.03 & & \\
\hline
Type-P7 & 14.11 & Type-Q2 & 26.24 & Type-R2 & 14.17 &  & \\
\hline
\end{tabular}
\caption{Estimated values of $m_{\beta\beta}$ associated with textures under EP.}
\label{table:Values of DBB under EP}
\end{table}

\begin{table}[!]
\centering
\begin{tabular}{P{2cm}P{.5cm}P{.5cm}P{.5cm}P{.5cm}P{.5cm}P{.5cm}P{.5cm}P{.5cm}P{.5cm}P{.5cm}} 
\hline\hline
Fields & $D_{l_L}$ & $e_R$ & $\mu_R$ & $\tau_R$ & $\Phi$ & $\nu_R$ & $\Delta$ & $\eta$ & $\kappa$ & $\xi$ \\ 
\hline\hline
$SU(2)_L$ & $2$ & $1$ & $1$ & $1$ & $2$ & $1$ & $3$ & $1$ & $1$ & $1$\\
\hline
$A_4$ & $3$ & $1$ & $1''$ & $1'$ & $3$ & $3$ & $3$ & $1$ & $3$ & $3$\\ 
\hline
$Z_2$ & $1$ & $-1$ & $-1$ & $-1$ & $-1$ & $-1$ & $1$ & $1$ & $1$ & $1$\\
\hline
\end{tabular}
\caption{Transformation properties of the various fields under $SU(2)_L \times A_4 \times Z_2$ group.}
\label{Field Chart of Type-A3}
\end{table}

\begin{table}[!]
\centering
\begin{tabular}{P{2cm}P{.5cm}P{.5cm}P{.5cm}P{.5cm}P{.5cm}P{.5cm}P{.5cm}P{.5cm}}
\hline\hline
Fields & $D_{l_L}$ & $e_R$ & $\mu_R$ & $\tau_R$ & $\Phi$ & $\nu_R$ & $\eta$ & $\Delta$  \\ 
\hline\hline
$SU(2)_L$ & $2$ & $1$ & $1$ & $1$ & $2$ & $1$ & $1$ & $3$  \\
\hline
$A_4$ & $3$ & $1$ & $1''$ & $1'$ & $3$ & $3$ & $1$ & $3$ \\ 
\hline
\end{tabular}
\caption{Transformation properties of the various fields under $SU(2)_L \times A_4$ group.}
\label{Field Chart of Type-R4}
\end{table}

\begin{table*}[!]
\centering
\begin{tabular}{P{1.5cm}P{0.5cm}P{2.5cm}P{2.5cm}P{0.3cm}P{0.3cm}P{0.3cm}} 
\hline\hline
Fields & $D_{l_{L}}$ & $l_{R}$ & $\nu_{l_R}$ & $H$ & $\kappa$ & $\Delta$ \\ 
\hline\hline
$SU(2)_{L}$ & 2 & 1 & 1 & 2 & 1 & 3 \\
\hline
$\Delta(27)$ & 3 & $(1_{20},1_{00},1_{10})$ & $(1_{00},1_{20},1_{10})$ & $3^*$ & $1_{10}$ & 3 \\
\hline
\end{tabular}
\caption{ The transformation properties of various fields under $SU(2)_L \times \Delta(27)$.} 
\label{Field Chart of Type-B8}
\end{table*}

\begin{table*}[!]
\centering
\begin{tabular}{P{1.5cm}P{0.5cm}P{2.5cm}P{2.5cm}P{0.3cm}P{0.3cm}P{0.3cm}P{0.3cm}} 
\hline\hline
Fields & $D_{l_{L}}$ & $l_{R}$ & $\nu_{l_R}$ & $\Phi$ & $\kappa$ & $\chi$ & $\Delta$ \\ 
\hline\hline
$SU(2)_{L}$ & 2 & 1 & 1 & 2 & 1 & 1 & 3 \\
\hline
$T_7$ & 3 & $(1_{0},1_{1},1_{2})$ & $\bar{3}$ & $\bar{3}$ & $\bar{3}$ & $3$ & 3 \\
\hline
\end{tabular}
\caption{ The transformation properties of various fields under $SU(2)_L \times T_7$.} 
\label{Field Chart of Type-P4}
\end{table*}

\FloatBarrier

\section{Product Rules of $A_4$ \label{appendix b}}

The discrete group $A_4$ is a subgroup of $SU(3)$ and it has 12 elements. The group has four irreducible representation: three one dimensional representations and one of dimension three. Under $A_4$, the product of two triplets gives,

\begin{equation}
3\times3=1+1'+1''+3_S+3_A,
\end{equation}

where,

\begin{eqnarray}
1&=&(a_1 b_1+a_2 b_2+a_3 b_3)\\
1'&=&(a_1 b_1+\omega^2 a_2b_2+\omega a_3b_3)\\
1''&=&(a_1b_1+\omega a_2 b_2+ \omega^2 a_3b_3)\\
(3\times3)_S&=&\begin{bmatrix}
a_2 b_3+a_3 b_2\\
a_3 b_1+a_1 b_3\\
a_1 b_2+a_2 b_1\\
\end{bmatrix},\\
(3\times3)_A&=&\begin{bmatrix}
a_2 b_3-a_3 b_2\\
a_3 b_1-a_1 b_3\\
a_1 b_2-a_2 b_1\\
\end{bmatrix}.
\end{eqnarray}

The trivial singlet can be obtained from the following singlet product rules,

\begin{equation}
1\times1=1,\,\,1'\times1''=1,\,\,1''\times1'=1.
\end{equation}

\section{Product Rules of $\Delta(27)$ \label{appendix a}}

The non-Abelian discrete group $\Delta\,(27)$ is a subgroup of $SU(3)$ has 27 elements which can be divided into 11 equivalence classes \cite{Luhn:2007uq}. The group has two three dimensional representations and nine one-dimensional representations.

The group multiplication rules are shown below,

\begin{eqnarray}
3 \times 3&=&3^*_{S_1}+3^*_{S_2}+3^*_{A},\\
3 \times 3^* &=& \sum_{r=0}^{2} 1_{r,0}+\sum_{r=0}^{2} 1_{r,1} \nonumber\\&& +\sum_{r=0}^{2} 1_{r,2},\\
1_{r,p}\times 1_{r', p'}&=&1_{(r+r') \,\text{mod}\,3,\,\, (p+p') \,\text{mod}\, 3},\\
\end{eqnarray}

where,

\begin{eqnarray}
(3\times3)_{3^*_{S_1}}&=&\begin{bmatrix}
a_1 b_1 \\
a_2 b_2\\
a_3 b_3\\
\end{bmatrix},\\
(3\times3)_{3^*_{S_2}}&=&\frac{1}{2}\begin{bmatrix}
a_2 b_3+a_3 b_2\\
a_3 b_1+a_1 b_3\\
a_1 b_2+a_2 b_1\\
\end{bmatrix},\\
(3\times3)_{3^*_{A}}&=&\frac{1}{2}\begin{bmatrix}
a_2 b_3-a_3 b_2\\
a_3 b_1-a_1 b_3\\
a_1 b_2-a_2 b_1\\
\end{bmatrix},
\end{eqnarray}

and,

\begin{eqnarray}
1_{00}&=&(a_1\bar{b_1}+a_2\bar{b_2}+a_3\bar{b_3})\\
1_{10}&=&(a_1\bar{b_1}+\omega^2 a_2\bar{b_2}+\omega a_3\bar{b_3})\\
1_{20}&=&(a_1\bar{b_1}+\omega a_2\bar{b_2}+\omega^2 a_3\bar{b_3})\\
1_{01}&=&(a_1\bar{b_2}+ a_2\bar{b_3}+a_3\bar{b_1})\\
1_{11}&=&(a_1\bar{b_2}+\omega^2 a_2\bar{b_3}+\omega a_3\bar{b_1})\\
1_{21}&=&(a_1\bar{b_2}+\omega a_2\bar{b_3}+\omega^2 a_3\bar{b_1})\\
1_{02}&=&(a_1\bar{b_3}+ a_2\bar{b_1}+a_3\bar{b_2})\\
1_{12}&=&(a_1\bar{b_3}+ \omega^2 a_2\bar{b_1}+\omega a_3\bar{b_2})\\
1_{22}&=&(a_1\bar{b_3}+ \omega a_2\bar{b_1}+\omega^2 a_3\bar{b_2})
\end{eqnarray}

\section{Product Rules of $T_7$ \label{appendix c}}

The group $T_7$ has five irreducible representation: three one dimensional representations, two of dimension three. Under this group, the product of two triplets gives,

\begin{eqnarray}
3 \times 3&=&\begin{bmatrix}
x_3 \bar{y_3}\\
x_1 \bar{y_1}\\
x_2 \bar{y_2}\\
\end{bmatrix}_3+\begin{bmatrix}
x_2 \bar{y_3}\\
x_3 \bar{y_1}\\
x_1 \bar{y_2}\\
\end{bmatrix}_{\bar{3}}+\begin{bmatrix}
x_3 \bar{y_2}\\
x_1 \bar{y_3}\\
x_2 \bar{y_1}\\
\end{bmatrix}_{\bar{3}},\\
3 \times \bar{3} &=& \sum_{a=0}^{2} 1_a+\begin{bmatrix}
x_2 \bar{y_1}\\
x_3 \bar{y_2}\\
x_1 \bar{y_3}\\
\end{bmatrix}_3+\begin{bmatrix}
x_1 \bar{y_2}\\
x_2 \bar{y_3}\\
x_3 \bar{y_1}\\
\end{bmatrix}_{\bar{3}},
\end{eqnarray}

where,

\begin{eqnarray}
\sum_{a=0}^{2} 1_a&=& (x_1\bar{y_1}+\omega^{2a}x_2\bar{y_2}+\omega^{a}x_3\bar{y_3})\\
\end{eqnarray}

Also, 

\begin{equation}
1_2\times1_1=1_0,\,\,1_1\times1_2=1_0,\,\,1_0\times1_0=1_0,
\end{equation}

The details of $T_7$ group can be found in ref \cite{Luhn:2007uq}.


\begin{thebibliography}{99}

\bibitem{a}
S. L. Glashow, \emph{Partial Symmetries of Weak Interactions}, \emph{Nucl. Phys.} {\bf 22} (1961) 579-588.

\bibitem{b}
S. Weinberg, \emph{A Model of Leptons}, \emph{Phys. Rev. Lett.} {\bf 19} (1967) 1264–1266.

\bibitem{c}
M. Herrero, \emph{The Standard model}, \emph{NATO Sci. Ser. C} {\bf 534} (1999) 1-59.

\bibitem{d}
V. A. Bednyakov, N. D. Giokaris, A. V. Bednyakov, \emph{On Higgs mass generation mechanism in the Standard Model}, \emph{Phys. Part. Nucl.} {\bf 39} (2008) 13–36. 

\bibitem{e}
S. D. Bass, A. De Roeck, M. Kado, \emph{The Higgs boson implications and prospects for future
discoveries}, \emph{Nature Rev. Phys.} {\bf 3(9)} (2021) 608.


\bibitem{f1}
B. Pontecorvo, \emph{Mesonium and anti-mesonium}, \emph{ Sov. Phys. JETP} {\bf 6} (1957) 429.


\bibitem{f}
Y. Farzan, M. Tortola, \emph{Neutrino oscillations and Non-Standard Interactions}, \emph{Front. in Phys.} {\bf 6} (2018) 10.

\bibitem{g}
S. A. R. Ellis, K. J. Kelly, S. W. Li, \emph{Current and Future Neutrino Oscillation Constraints on
Leptonic Unitarity}, \emph{JHEP} {\bf 12} (2020) 68.

\bibitem{h}
G. Fantini, A. Gallo Rosso, F. Vissani, V. Zema, \emph{Introduction to the Formalism of Neutrino
Oscillations}, \emph{ Adv. Ser. Direct. High Energy Phys.} {\bf 28} (2018) 37–119. 


\bibitem{i}
H. Arodź, \emph{Relativistic Quantum Mechanics of the Majorana Particle}, \emph{Acta Phys. Polon. B} {\bf 50} (2019) 2165–2187.


\bibitem{j}
L. Borsten, M. J. Duff, \emph{Majorana Fermions in Particle Physics, Solid State and Quantum Information}, \emph{Subnucl. Ser.} {\bf 53} (2017) 77-121.

\bibitem{k}
B. Adhikary, M. Chakraborty, A. Ghosal, \emph{ Masses, mixing angles and phases of general Majorana
neutrino mass matrix}, \emph{JHEP} {\bf 10} (2013) 043.


\bibitem{k1}
Z.-z. Xing, \emph{Flavor structures of charged fermions and massive neutrinos}, \emph{Phys. Rept.} {\bf 854} (2020) 1–147.


\bibitem{r1}
A. Loureiro, et al., \emph{On The Upper Bound of Neutrino Masses from Combined Cosmological
Observations and Particle Physics Experiments}, \emph{ Phys. Rev. Lett.} {\bf 123} (2019) 081301.


\bibitem{r2}
A. Upadhye, \emph{Neutrino mass and dark energy constraints from redshift-space distortions}, \emph{JCAP} {\bf 05} (2019) 041.

\bibitem{r3}
S. Roy Choudhury, S. Hannestad, \emph{Updated results on neutrino mass and mass hierarchy from cosmology with Planck 2018 likelihoods}, \emph{JCAP} {\bf 07} (2020) 037.

\bibitem{r4}
C. Yèche, N. Palanque-Delabrouille, J. Baur, H. du Mas des Bourboux, \emph{Constraints on neutrino
masses from Lyman-alpha forest power spectrum with BOSS and XQ-100}, \emph{JCAP} {\bf 06} (2017) 047.

\bibitem{Maki:1962mu}
Z.~Maki, M.~Nakagawa, S.~Sakata, {Remarks on the unified model of elementary
  particles}, Prog. Theor. Phys. 28 (1962) 870--880.
\newblock \href {https://doi.org/10.1143/PTP.28.870}
  {\path{doi:10.1143/PTP.28.870}}.



\bibitem{l1}
W. Grimus, L. Lavoura, \emph{The Seesaw mechanism at arbitrary order: Disentangling the small scale
from the large scale}, \emph{JHEP} {\bf 11} (2000) 042.

\bibitem{l2}
E. K. Akhmedov, G. C. Branco, M. N. Rebelo, \emph{Seesaw mechanism and structure of neutrino
mass matrix}, \emph{Phys. Lett. B} {\bf 478} (2020) 215–223.


\bibitem{l3}
O. G. Miranda, J. W. F. Valle, \emph{Neutrino oscillations and the seesaw origin of neutrino mass}, \emph{Nucl.
Phys. B} {\bf 908} (2016) 436–455.


\bibitem{m}
P. A. Zyla, et al., \emph{Review of Particle Physics}, \emph{PTEP} {\bf 2020} (2020) 083C01.

\bibitem{n}
I. Esteban, M. C. Gonzalez-Garcia, M. Maltoni, T. Schwetz, A. Zhou, \emph{The fate of hints: updated
global analysis of three-flavor neutrino oscillations}, \emph{JHEP} {\bf 09} (2020) 178.

\bibitem{o1}
 M. J. Dolinski, A. W. P. Poon, W. Rodejohann, \emph{Neutrinoless Double-Beta Decay: Status and
Prospects}, \emph{Ann. Rev. Nucl. Part. Sci.} {\bf 69} (2019) 219–251.

\bibitem{o2}
 J. D. Vergados, H. Ejiri, F. Šimkovic, \emph{Neutrinoless double beta decay and neutrino mass}, \emph{Int. J.
Mod. Phys. E} {\bf 25} (2016) 1630007.


\bibitem{o3}
  H. Päs, W. Rodejohann, \emph{Neutrinoless Double Beta Decay}, \emph{New J. Phys.} {\bf 17} (2015) 115010.
  

\bibitem{o4}
  F. Simkovic, S. M. Bilenky, A. Faessler, T. Gutsche, \emph{Possibility of measuring the CP Majorana
phases in $0\nu\beta\beta$ decay}, \emph{Phys. Rev. D} {\bf 87} (2013) 073002.

\bibitem{p}
  H.~Ejiri, \emph{Neutrino-mass
  sensitivity and nuclear matrix element for neutrinoless double beta decay
}, \emph{Universe} {\bf 6} (2020) 225.

\bibitem{Agostini:2022zub}
M.~Agostini, G.~Benato, J.~A. Detwiler, J.~Men\'endez, F.~Vissani, {Toward the
  discovery of matter creation with neutrinoless
  \ensuremath{\beta}\ensuremath{\beta} decay}, Rev. Mod. Phys. 95~(2) (2023)
  025002.
\newblock \href {http://arxiv.org/abs/2202.01787} {\path{arXiv:2202.01787}},
  \href {https://doi.org/10.1103/RevModPhys.95.025002}
  {\path{doi:10.1103/RevModPhys.95.025002}}.




\bibitem{CUORE:2019yfd}
D.~Q. Adams et~al.
\newblock {Improved Limit on Neutrinoless Double-Beta Decay in $^{130}$Te with
  CUORE}.
\newblock {\em Phys. Rev. Lett.}, 124(12):122501, 2020.

\bibitem{GERDA:2019ivs}
M.~Agostini et~al.
\newblock {Probing Majorana neutrinos with double-$\beta$ decay}.
\newblock {\em Science}, 365:1445, 2019.

\bibitem{KamLAND-Zen:2016pfg}
A.~Gando et~al.
\newblock {Search for Majorana Neutrinos near the Inverted Mass Hierarchy
  Region with KamLAND-Zen}.
\newblock {\em Phys. Rev. Lett.}, 117(8):082503, 2016.
\newblock [Addendum: Phys.Rev.Lett. 117, 109903 (2016)].


\bibitem{p1}
  R. Arnold, et al., \emph{Measurement of the distribution of $^{207}$Bi depositions on calibration sources
for SuperNEMO}, \emph{JINST} {\bf 16} (2021) T07012.



\bibitem{p2}
 A. S. Barabash, et al., \emph{Calorimeter development for the SuperNEMO double beta decay experiment}, \emph{Nucl. Instrum. Meth. A} {\bf 868} (2017) 98–108.

\bibitem{p3}
D. Q. Adams, et al., \emph{Improved Limit on Neutrinoless Double-Beta Decay in 130Te with CUORE}, \emph{Phys. Rev. Lett.} {\bf 124} (2020) 122501.
  
\bibitem{p4}
E. Andreotti, et al., \emph{Double-beta decay of 130Te to the first 0+ excited state of 130Xe with
CUORICINO}, \emph{Phys. Rev. C} {\bf 85} (2012) 045503.




\bibitem{Ludl:2014axa}
P.~O. Ludl, W.~Grimus, {A complete survey of texture zeros in the lepton mass
  matrices}, JHEP 07 (2014) 090, [Erratum: JHEP 10, 126 (2014)].
\newblock \href {http://arxiv.org/abs/1406.3546} {\path{arXiv:1406.3546}},
  \href {https://doi.org/10.1007/JHEP07(2014)090}
  {\path{doi:10.1007/JHEP07(2014)090}}.



\bibitem{Singh:2018bap}
M.~Singh, {Investigating the hybrid textures of neutrino mass matrix for near
  maximal atmospheric neutrino mixing}, Adv. High Energy Phys. 2018 (2018)
  5806743.
\newblock \href {http://arxiv.org/abs/1803.10754} {\path{arXiv:1803.10754}},
  \href {https://doi.org/10.1155/2018/5806743}
  {\path{doi:10.1155/2018/5806743}}.

\bibitem{Lashin:2009yd}
E.~I. Lashin, N.~Chamoun, {One vanishing minor in the neutrino mass matrix},
  Phys. Rev. D 80 (2009) 093004.
\newblock \href {http://arxiv.org/abs/0909.2669} {\path{arXiv:0909.2669}},
  \href {https://doi.org/10.1103/PhysRevD.80.093004}
  {\path{doi:10.1103/PhysRevD.80.093004}}.

\bibitem{Planck:2018vyg}
N.~Aghanim, et~al., {Planck 2018 results. VI. Cosmological parameters}, Astron.
  Astrophys. 641 (2020) A6, [Erratum: Astron.Astrophys. 652, C4 (2021)].
\newblock \href {http://arxiv.org/abs/1807.06209} {\path{arXiv:1807.06209}},
  \href {https://doi.org/10.1051/0004-6361/201833910}
  {\path{doi:10.1051/0004-6361/201833910}}.


\bibitem{q}
Y. Cai, J. Herrero-García, M. A. Schmidt, A. Vicente, R. R. Volkas, \emph{From the trees to the forest:
a review of radiative neutrino mass models}, \emph{Front. in Phys.} {\bf 5} (2017) 63.


\bibitem{r}
M. C. Gonzalez-Garcia, M. Maltoni, T. Schwetz, \emph{NuFIT: Three-Flavour Global Analyses of
Neutrino Oscillation Experiments}, \emph{Universe} {\bf 7} (2021) 459.

\bibitem{deGouvea:2008nm}
A.~de~Gouvea, J.~Jenkins, {The Physical Range of Majorana Neutrino Mixing
  Parameters}, Phys. Rev. D 78 (2008) 053003.
\newblock \href {http://arxiv.org/abs/0804.3627} {\path{arXiv:0804.3627}},
  \href {https://doi.org/10.1103/PhysRevD.78.053003}
  {\path{doi:10.1103/PhysRevD.78.053003}}.


\bibitem{t}
Soumita Pramanick, Amitava Raychaudhuri, \emph{Three-Higgs-doublet model under $A_4$ symmetry implies alignment}, \emph{JHEP} {\bf 01} (2018) 011.

\bibitem{Branco:1983tn}
G.~C. Branco, J.~M. Gerard, W.~Grimus, {GEOMETRICAL T VIOLATION}, Phys. Lett. B
  136 (1984) 383--386.
\newblock \href {https://doi.org/10.1016/0370-2693(84)92024-0}
  {\path{doi:10.1016/0370-2693(84)92024-0}}.

\bibitem{Dey:2022qpu}
M.~Dey, P.~Chakraborty, S.~Roy, {The \ensuremath{\mu}-\ensuremath{\tau} mixed
  symmetry and neutrino mass matrix}, Phys. Lett. B 839 (2023) 137767.
\newblock \href {http://arxiv.org/abs/2211.01314} {\path{arXiv:2211.01314}},
  \href {https://doi.org/10.1016/j.physletb.2023.137767}
  {\path{doi:10.1016/j.physletb.2023.137767}}.


\bibitem{s1}
Ernest Ma, \emph{A(4) symmetry and neutrinos with very different masses}, \emph{Universe} {\bf 70} (2004) 031901.

\bibitem{s2}
Ernest Ma, Daniel Wegman, \emph{Nonzero $\theta_{13}$ for Neutrino Mixing in the Context of $A_4$ Symmetry}, \emph{Phys. Rev. Lett.} {\bf 107} (2011) 061803.

\bibitem{s3}
 S. Morisi, E. Peinado, Y. Shimizu, and J. W. F. Valle, \emph{Relating quarks and leptons with the $T_7$ flavour group}, \emph{Phys. Lett. B} {\bf 742} (2015) 99--106.

\bibitem{Luhn:2007uq}
C.~Luhn, S.~Nasri, P.~Ramond, {The Flavor group Delta(3n**2)}, J. Math. Phys.
  48 (2007) 073501.
\newblock \href {http://arxiv.org/abs/hep-th/0701188}
  {\path{arXiv:hep-th/0701188}}, \href {https://doi.org/10.1063/1.2734865}
  {\path{doi:10.1063/1.2734865}}.


\end{thebibliography}
\end{document}